\documentclass[onecolumn]{aastex631}
\renewcommand{\vec}[1]{\mathbf{#1}}

\usepackage{hyperref}
\usepackage[all]{hypcap}

\begin{document}

\title{Efficient numerical treatment of ambipolar and Hall drift as hyperbolic system}
\shorttitle{Hyperbolic ambipolar and Hall drift}

\author{M. Rempel}
\affiliation{High Altitude Observatory, NCAR, P.O. Box 3000, Boulder, Colorado 80307, USA}

\author{D. Przybylski}
\affiliation{Max Planck Institute for Solar System Research, Justus-von-Liebig-Weg 3, 37077 G{\"o}ttingen, Germany}

\shortauthors{Rempel et al.}

\email{rempel@ucar.edu}

\begin{abstract}
Partially ionized plasmas, such as the solar chromosphere, require a generalized Ohm's law including the effects of ambipolar and Hall drift. While both describe transport processes that arise from the multifluid equations and are therefore of hyperbolic nature, they are often incorporated in models as a diffusive, i.e. parabolic process. While the formulation as such is easy to include in standard MHD models, the resulting diffusive time-step constraints do require often a computationally more expensive implicit treatment or super-time-stepping approaches. In this paper we discuss an implementation that retains the hyperbolic nature and allows for an explicit integration with small computational overhead. In the case of ambipolar drift, this formulation arises naturally by simply retaining a time derivative of the drift velocity that is typically omitted. This alone leads to time-step constraints that are comparable to the 
native MHD time-step constraint for a solar setup including the region from photosphere to lower solar corona. We discuss an accelerated treatment that can further reduce time-step constraints if necessary. In the case of Hall drift we propose a hyperbolic formulation that is numerically similar to that for the ambipolar drift and we show that the combination of both can be applied to simulations of the solar chromosphere at minimal computational expense.    
\end{abstract}


\received{}
\accepted{}

\section{Introduction}
In plasmas with a low degree of ionization and a sufficiently low ion-neutral collision rate the neutral and ionized fluids are only weakly coupled, which can result in substantial drift velocities between the ionized and neutral components. This process is properly modeled as a hyperbolic multifluid MHD problem in which only the ionized component is coupled to the magnetic field through the induction equation and Lorentz force in the momentum equation. As described in detail in \citet{Braginskii:1965} the problem can be reduced to a single fluid treatment with a generalized Ohm's law including a ion-neural drift velocity proportional to the Lorentz force in situations where the time-scale of evolution is substantially longer than the collision time-scale between ions and neutrals. Under such conditions the ion-neutral drift can be modeled as a parabolic diffusion problem, hence the description as ``ambipolar diffusion". The resulting ambipolar diffusivity is quadratic in the magnetic field strength and has the tendency to create sharp current sheets around null points \citep{Brandenbrug:Zweibel:1994:ambdiff}. In addition to ambipolar diffusion the generalized Ohm's law includes the Hall term, which results from the Lorentz force in the electron momentum equation. Unlike ambipolar diffusion, the Hall effect is nondissipative (electric field perpendicular to the current), but has formally also a parabolic form. The Hall term can play a critical role in reconnection and does affect reconnection rates \citep[see, e.g.][]{Birn:2001:GEMChallenge,Shi:2019:HallReconnection}. 

While such treatment as the parabolic diffusion problem is conceptually simple, it can be numerically challenging if the resulting diffusive time-step constraint is severe, which is the case when ambipolar and Hall diffusion are applied in the solar chromosphere. For this reason recent implementations such as \cite{Osullivan:2007:STS,Gonzalez:2018:ManchaSTS,Nobrega:2020:BifrostSTS} use a super-time-stepping (STS) scheme \citep{alexiades1996super} to overcome these constraints. STS has been also applied to heat conduction problems by \citet{Meyer:2012:STS,Iijima2015super}. In STS the number of required substeps scales asymptotically as $n^2\sim {\Delta t}_{\rm MHD}/{\Delta t}_{\rm diff}$, where ${\Delta t}_{\rm MHD}$ and ${\Delta t}_{\rm diff}$ are the MHD and (explicit) diffusive time-step limits, respectively (we assume here that ${\Delta t}_{\rm diff} < {\Delta t}_{\rm MHD}$). While this provides a substantial improvement over explicit time stepping with $n\sim {\Delta t}_{\rm MHD}/{\Delta t}_{\rm diff}$, the computational cost can be still substantial when ${\Delta t}_{\rm diff} \ll {\Delta t}_{\rm MHD}$. In the case of heat conduction \citet{Rempel:2017:corona} used a different approach by casting the parabolic diffusion problem again as a hyperbolic transport problem and solving the resulting damped wave equation explicitly. It was shown that such explicit integration can be conducted with ${\Delta t}_{\rm MHD}$ by limiting the hyperbolic transport velocity accordingly. This approach has been implemented in other codes used for simulations of the solar corona \citep[e.g.][]{fan2017streamer,warnecke:2020:conduction}. 

In this paper we explore a similar approach for the treatment of ambipolar diffusion, which we implemented in the MURaM radiative MHD code \citep{Voegler:2005,Rempel:2017:corona}. In Section \ref{sec:ambdiff} we go through the derivation of the equations following \citet{Pandey:Wardle:2008} and highlight which terms need to be kept to arrive naturally at a hyperbolic set of equations for ambipolar drift as presented in Section \ref{sec:hyperbolic}. It turns out that the resulting hyperbolic set of equations is in most cases already more suitable for explicit numerical time integration than the parabolic one. In Section \ref{sec:hyperbolic_acc} we show how the time-step constraint can be further relaxed by adjusting the ion-neutral collision frequency while maintaining the correct value for the resulting ambipolar drift velocity. The expression for the ambipolar heating term that is consistent with the hyperbolic treatment is given in Section \ref{sec:ambipolar_heating}. In Section \ref{sec:amb_hall} we introduce a generalization that also includes the Hall effect. Details specific to the implementation in the MURaM code are given in Section \ref{sec:numerical}. In Section \ref{sec:tabs} we provide the ion-neutral collision frequencies that are used in the MURaM code, and finally Section \ref{sec:tests} provides numerical tests in 1D and 2D that demonstrate utility and limitations of the hyperbolic treatment. Conclusions are given in Section \ref{sec:concl}.

\section{Ambipolar and Hall drift}
\label{sec:ambdiff}
Formally the ambipolar diffusion approximation follows from multifluid MHD after formulating the problem in terms of the drift velocity between ions and neutrals and simplifying the evolution equation for the drift velocity, which is detailed in \citet{Pandey:Wardle:2008}; we summarize here the key expressions for reference. We start with the momentum equations for the electron, ion, and neutral fluid:
\begin{eqnarray}
	\varrho_e\frac{d \vec{v}_e}{dt}&=&-\nabla p_e +n_e q_e \left(\vec{E}+\frac{\vec{v_e}}{c}\times\vec{B}\right)+\varrho_e\vec{g}-\varrho_e \nu_{ei}(\vec{v}_e-\vec{v}_i)-\varrho_e \nu_{en}(\vec{v}_e-\vec{v}_n),\\
	\varrho_i\frac{d \vec{v}_i}{dt}&=&-\nabla p_i +n_i q_i \left(\vec{E}+\frac{\vec{v_i}}{c}\times\vec{B}\right)+\varrho_i\vec{g}-\varrho_i \nu_{ie}(\vec{v}_i-\vec{v}_e)-\varrho_i \nu_{in}(\vec{v}_i-\vec{v}_n),\\
	\varrho_n\frac{d \vec{v}_n}{dt}&=&-\nabla p_n +\varrho_n\vec{g}-\varrho_n \nu_{ne}(\vec{v}_n-\vec{v}_e)-\varrho_n \nu_{ni}(\vec{v}_n-\vec{v}_i).
\end{eqnarray}
Charge neutrality of the plasma implies $n_e q_e=-n_i q_i$, and the electric current is given by: $\vec{J}=n_e q_e \vec{v}_e+n_i q_i\vec{v}_i=n_e q_e (\vec{v}_e-\vec{v}_i)=n_e e (\vec{v}_i-\vec{v}_e)$, with $q_e=-e$. These equations are complemented  with the continuity equations for each component. Neglecting the contribution from the electron fluid since $\varrho_e\ll \varrho_i$, we can define the bulk flow components as
\begin{eqnarray}
	\varrho&=& \varrho_i+\varrho_n,\\	
	\vec{v}&=&(\varrho_i \vec{v}_i+\varrho_n\vec{v}_n)/\varrho = \vec{v}_i-D\vec{v}_D=\vec{v}_n+(1-D)\vec{v}_D,\label{def_v}
\end{eqnarray}
with
\begin{eqnarray}	
	D&=&\frac{\varrho_n}{\varrho},\\
	\vec{v}_D&=&\vec{v}_i-\vec{v}_n.
\end{eqnarray}
As shown in  \citet{Pandey:Wardle:2008}, the single fluid momentum equation can be derived by adding up the momentum equations of the electron, ion, and neutral fluid. The resulting equation does contain a term quadratic in $\vec{v}_D$, which can be neglected under most conditions:
\begin{equation}
	\frac{\partial \varrho \vec{v}}{\partial t}+\nabla\cdot\left(\varrho\vec{v}\vec{v}+\frac{\varrho_i\varrho_n}{\varrho}\vec{v}_D\vec{v}_D\right)=-\nabla p+\frac{1}{c}\vec{J}\times\vec{B}+\varrho\vec{g}.\label{bulk}
\end{equation}
The momentum equation for electrons results (after neglecting the electron inertia term) in the following expression for the electric field in the plasma reference frame moving with the bulk fluid velocity
$\vec{v}$:
\begin{equation}
   \vec{E}+\frac{\vec{v}}{c}\times\vec{B}=-\frac{\nabla p_e}{n_e e}+\frac{m_e}{n_e e^2}(\nu_{ei}+\nu_{in})\vec{J}+\frac{1}{n_e e c}\vec{J}\times\vec{B}-D\frac{\vec{v}_D}{c}\times\vec{B}-\frac{m_e}{e}\nu_{en}\vec{v}_D.\label{ohm}
\end{equation}
From the difference of the plasma (electron + ion) and neutrals momentum equation follows the following expression for the ion-neutral drift velocity:
\begin{equation}
	\frac{\varrho_i\varrho_n}{\varrho}\left(\frac{d \vec{v}_i}{dt}-\frac{d \vec{v}_n}{dt}\right)=-D\nabla p+\nabla p_n+\frac{D}{c}\vec{J}\times\vec{B}-\left(\varrho_i\nu_{in}+\varrho_e\nu_{en}\right)\vec{v}_D+\frac{m_e}{e}\nu_{en}\vec{J}.\label{drift}
\end{equation}
Eqs. (\ref{bulk}-\ref{drift}) are exact with no additional assumptions besides neglecting the electron inertia. In the following we use a simplified set of equations and refer to \citet{Pandey:Wardle:2008} for a detailed discussion under which conditions these simplifications are justified. We neglect the last terms on the rhs of Eqs. (\ref{ohm}) and (\ref{drift}), the pressure contributions in Eqs. (\ref{ohm}) and (\ref{drift}) as well as ohmic dissipation (second term on rhs of Eq. (\ref{ohm})). Since we focus the discussion in following sections first on the treatment of ambipolar diffusion, we also neglect the Hall term (third term on rhs of Eq. (\ref{ohm})), but we will discuss ways to include it in Section \ref{sec:amb_hall}. With these simplifications Eqs. (\ref{ohm}) and (\ref{drift}) reduce to
\begin{eqnarray}
	\vec{E}+\frac{\vec{v}}{c}\times\vec{B}&=&-D\frac{\vec{v}_D}{c}\times\vec{B}\label{ohm2},\\
	\frac{\varrho_i\varrho_n}{\varrho}\left(\frac{d \vec{v}_i}{dt}-\frac{d \vec{v}_n}{dt}\right)&=&\frac{D}{c}\vec{J}\times\vec{B}-\varrho_i\nu_{in}\vec{v}_D.\label{drift2}
\end{eqnarray}	
The standard treatment  of the ion-neutral drift in terms of an ambipolar diffusion term follows from these equations after neglecting the acceleration terms on the l.h.s. of Eq. (\ref{drift2}), leading to
\begin{equation}
	\vec{v}_D=\frac{D}{\varrho_i\nu_{in}c}\vec{J}\times\vec{B}\label{drift_diffusion}
\end{equation}	
and an induction equation of the form
\begin{equation}
	\frac{\partial\vec{B}}{\partial t}=-c\nabla\times\vec{E}=-\nabla\times\left(\vec{v}\times\vec{B}+\frac{D^2}{\varrho_i\nu_{in}c}(\vec{J}\times\vec{B})\times\vec{B}\right).
\end{equation}
This equation is of parabolic nature due to the presence of the ambipolar diffusion term with a diffusivity of
\begin{equation}
	D_{\rm Amb}=\frac{D^2 \vert \vec{B}\vert^2}{4\pi \varrho_i\nu_{in}}\label{ambdiff}\;.
\end{equation}
Consequently, an explicit integration of the induction equation imposes strict (parabolic) time-step constraints of
\begin{equation}
	\Delta t_{\rm Amb}^{\rm par} \sim \frac{\Delta x ^2}{D_{\rm Amb}}\sim \frac{\Delta x ^2}{\vert \vec{B}\vert^2},\label{dt_parabolic}
\end{equation}	
which are particularly severe in simulations that have high resolution and strong magnetic field, since the time-step constraints of the ideal MHD system only scales as $\Delta x/\vert \vec{B}\vert$. For typical applications to the solar chromosphere, which we will discuss in Section \ref{sec:test_2D}, the explicit ambipolar time-step constraints can be orders of magnitude more severe than the MHD time-step constraints. Consequently, ambipolar diffusion has been treated in codes through either implicit or super-time-stepping methods \citep{Osullivan:2007:STS,Gonzalez:2018:ManchaSTS,Nobrega:2020:BifrostSTS} in order to avoid these limitations.

\subsection{Hyperbolic treatment}
\label{sec:hyperbolic}
Physically ambipolar diffusion can be also interpreted as a transport process with a velocity $\vec{v}_D$, which can be captured with a time step
\begin{equation}
	\Delta t_{\rm Drift} \sim \frac{\Delta x }{\vert D\vec{v}_D\vert}\sim \Delta t_{\rm Amb}^{\rm par} \frac{\vert\vec{B}\vert}{\vert \nabla\times\vec{B}\vert \Delta x}\;.
\end{equation}
While $\Delta t_{\rm Drift}$ is sufficient to capture the physical transport of magnetic field, using $\Delta t_{\rm Drift}$ in an explicit code is in general unstable as it violates the formal time-step constrained stemming from the parabolic nature of the underlying equations. Note that $\Delta t_{\rm Drift}$ can be substantially larger than $\Delta t_{\rm Amb}^{\rm par}$ in regions where the magnetic field is strong, while currents
are weak.

In order to avoid the rather stringent time-step constraints from the parabolic system, we investigate here a hyperbolic treatment that follows naturally from Eq. (\ref{drift2}) if we do not neglect the time derivative on the
left-hand side. Using Eq. (\ref{def_v}) we can write
\begin{eqnarray}
	\frac{d \vec{v}_i}{dt}-\frac{d \vec{v}_n}{dt}&=&\frac{\partial \vec{v}_D}{\partial t}+(\vec{v}_i\cdot\nabla)\vec{v}_i-(\vec{v}_n\cdot\nabla)\vec{v}_n\\\nonumber
		&=&\frac{\partial \vec{v}_D}{\partial t}+(\vec{v}\cdot\nabla)\vec{v}_D+(\vec{v}_D\cdot\nabla)\vec{v}+(2D-1)(\vec{v}_D\cdot\nabla)\vec{v}_D+\vec{v}_D(\vec{v}_D\cdot\nabla D).
\end{eqnarray}
The additional advective terms are small under most circumstances. Comparing them to the contribution from  $\nu_{in}\vec{v}_D$ (we assume here $D\sim 1$), their relative amplitude is on the order of $\mbox{max}(\vert\vec{v}\vert,\vert\vec{v}_D\vert)/(\Delta x \;\nu_{in})$. For conditions that are typical for chromospheric simulations ($v, v_D\sim$ 10 km s$^{-1}$, $\Delta x\sim 10$ km, $\nu_{in}\sim 10^4$ Hz (see Figure \ref{fig:1})) this ratio is on the order of $10^{-4}$. In the following we neglect these terms and focus only on the partial time derivative, which is the key term that is required in order to maintain the hyperbolic character of the system:
\begin{eqnarray}
	\frac{\partial \vec{v}_D}{\partial t}&=&\frac{\nu_{in}}{D}\left(\frac{D}{c\varrho_i\nu_{in}}\vec{J}\times\vec{B}-\vec{v}_D\right),\label{drift3}\\
	\frac{\partial\vec{B}}{\partial t}&=&-\nabla\times\left(\vec{v}\times\vec{B}+D\vec{v}_D\times\vec{B}\right).
\end{eqnarray}
Considering a simplified setup with $\vec{v}=0$, a background field $\vec{B}_0=B_0\vec{\hat{b}}$ with $B_0$ and $D$ not  varying in space, solutions obey the following damped wave equation:
\begin{equation}
	\frac{\partial^2 \vec{v}_D}{\partial t^2}+\frac{\nu_{in}}{D}\frac{\partial\vec{v}_D}{\partial t}+\frac{D B_0^2}{4\pi\varrho_i}\nabla\times\nabla\times\left[(\vec{v}_D\times\vec{\hat{b}})\times\vec{\hat{b}}\right]=0.\label{wave1}
\end{equation}
The maximum characteristic velocity is given by the Alv{\'e}n velocity of the ionized fraction of the plasma (with an additional factor of D): 
\begin{equation}
	C_{\rm Amb}^{\rm hyp}=\vert\vec{B}\vert\sqrt{\frac{D}{4\pi\varrho_i}}.\label{vai}
\end{equation} 
Unlike the parabolic time-step constraint given in Eq. (\ref{dt_parabolic}), the hyperbolic time-step constraint
\begin{equation}
	\Delta t_{\rm Amb}^{\rm hyp} \sim \frac{\Delta x}{C_{\rm Amb}}\sim \frac{\Delta x}{\vert \vec{B}\vert}\label{dt_hyperbolic}
\end{equation}	
is less severe in the limit of high resolution and strong field. Specifically the ratio of both is given by:
\begin{equation}
	\frac{\Delta t_{\rm Amb}^{\rm hyp}}{\Delta t_{\rm Amb}^{\rm par} }\sim D\frac{C_{\rm Amb}^{\rm hyp}}{\Delta x\,\nu_{in}}.
\end{equation}
In Section \ref{sec:test_2D} we will evaluate these quantities for a 2D chromospheric setting with a moderate vertical mean field strength of $25$~G. Typical values for $C_{\rm Amb}^{\rm hyp}$ are on the order of a several $1000$ km s$^{-1}$, which leads to $\Delta t_{\rm Amb}^{\rm hyp}$ being about $\sqrt{\varrho/\varrho_i}\sim 10-100$ times smaller than the MHD time-step constraint within the chromosphere (see Figure \ref{fig:1}). However, when considering a setup including a transition region, coronal values of the Alfv{\'e}n velocity will reach values comparable to $C_{\rm Amb}^{\rm hyp}$ in the chromosphere, which makes the hyperbolic treatment of the ambipolar drift not more time-step limiting than MHD itself for such a setup. 

\subsection{Hyperbolic treatment with modified collision frequencies}
\label{sec:hyperbolic_acc}
Here we suggest a minor modification of the Eq. (\ref{drift3}), which allows us to further relax time-step constraints without significantly impacting the accuracy of the physics. We introduce the concept here and refer to Section \ref{sec:tests} for application examples and comparison to the previous approaches. We rewrite Eq. (\ref{drift3}) as
\begin{equation}
	\frac{\partial \vec{v}_D}{\partial t}=\frac{1}{\tau}\left(\frac{D}{c\varrho_i\nu_{in}}\vec{J}\times\vec{B}-\vec{v}_D\right)\label{drift4}.
\end{equation}
Here we replaced the relaxation time-scale $D/\nu_{in}$ in front of the angular brackets by a generalized time-scale $\tau$, while we keep $\nu_{in}$ for the expression inside the angular brackets. Asymptotically $\vec{v}_D$ still relaxes toward $\frac{D}{c\varrho_i\nu_{in}}\vec{J}\times\vec{B}$, however, the relaxation time-scale is different. The corresponding damped wave equation Eq. (\ref{wave1}) is changed to
\begin{equation}
	\frac{\partial^2 \vec{v}_D}{\partial t^2}+\frac{1}{\tau}\frac{\partial\vec{v}_D}{\partial t}+\frac{D}{\nu_{in}\tau}\frac{D B_0^2}{4\pi\varrho_i}\nabla\times\nabla\times\left[(\vec{v}_D\times\vec{\hat{b}})\times\vec{\hat{b}}\right]=0.\label{wave2}
\end{equation}
This modification changes the maximum characteristic velocity of the hyperbolic ambipolar drift treatment to
\begin{equation}
	C_{\rm Amb}^{\rm hyp^*}=\vert\vec{B}\vert\sqrt{\frac{D}{4\pi\varrho_i}}\sqrt{\frac{D}{\nu_{in}\tau}}=\sqrt{\frac{D_{\rm Amb}}{\tau}}.
\end{equation}
Let us assume $C_{\rm MHD}$ is the maximum characteristic velocity of the MHD system, then a time-step constraint from ambipolar diffusion can be avoided by determining $\tau$ such that 
$ C_{\rm Amb}^{\rm hyp^*}<C_{\rm MHD}$, i.e.,
\begin{equation}
	\tau > \tau^*=\frac{D_{\rm Amb}}{C_{\rm MHD}^2}=\frac{D^2 \vert \vec{B}\vert^2}{4\pi \varrho_i\nu_{in}\; C_{\rm MHD}^2}.
\end{equation}
In order to maintain numerical stability some additional factors of order unity may be required, depending on the specifics of the implementation. In the MURaM code we use
\begin{equation}
		\tau=\mbox{max}\left(\frac{D}{\nu_{in}},f_{\rm N}\frac{D^2 \vert \vec{B}\vert^2}{4\pi \varrho_i\nu_{in}\; C_{\rm MHD}^2}\right),\label{tau_def}
\end{equation}	
where $f_{\rm N}$ is given by ($N$ is the dimensionality of the problem)
\begin{equation}
	f_{\rm N}=\mbox{min}(\Delta x_i^2)\sum_{i=1}^N\frac{1}{\Delta x_i^2}.\label{geo_fac}
\end{equation}

Furthermore Eq. (\ref{tau_def}) ensures that $\tau$ cannot drop below $\frac{D}{\nu_{in}}$ in regions where ambipolar diffusion does not limit the time step. In those regions $\tau$ will be substantially smaller than the numerical time step $\Delta t$ that follows from $C_{\rm MHD}$, i.e. Eq. (\ref{drift3}) needs to be integrated in a point-implicit manner for stability as further outlined in Section \ref{sec:numerical}. 

Formally the treatment of ambipolar diffusion introduced here is equivalent to the hyperbolic treatment of heat conduction in the MURaM code that is described in \citet{Rempel:2017:corona}.

\subsection{Heating from Ambipolar drift}
\label{sec:ambipolar_heating}
The total kinetic energy of the ion/neutral fluid can be written as
\begin{equation}
	\frac{1}{2}\varrho_i\vec{v}_i^2+\frac{1}{2}\varrho_n\vec{v}_n^2=\frac{1}{2}\varrho\vec{v}^2+\frac{1}{2}\frac{\varrho_i\varrho_n}{\varrho}\vec{v}_D^2.
\end{equation}

The ambipolar drift leads to an energy exchange between magnetic energy and the kinetic energy associated with the ion/neutral drift, whereas as the ion/neutral collisions lead to a dissipation
of that energy heating the plasma. From Eq. (\ref{drift3}) follows
\begin{equation}
	\frac{\varrho_i\varrho_n}{\varrho}\frac{\partial}{\partial t}\frac{\vec{v}_D^2}{2}=\frac{D_n}{c}\;\vec{v}_D\cdot (\vec{J}\times\vec{B})-\nu_{in}\varrho_i\vec{v}_D^2=W_{D}-Q_{\rm Amb}.\label{drift_energy}
\end{equation}
Here, $W_D$ given by
\begin{equation}
	W_D=\frac{D_n}{c}\;\vec{v}_D\cdot (\vec{J}\times\vec{B})\label{W_D}
\end{equation}
describes the energy exchange from magnetic to kinetic energy due to the ion-neutral drift, whereas	
\begin{equation}
	Q_{\rm Amb}=\nu_{in}\varrho_i\vec{v}_D^2\label{qamb_hyp}
\end{equation}
describes the energy exchange from kinetic energy to internal energy and this is the ambipolar heating term to be considered in the internal energy equation.
	
Under the assumption of ambipolar diffusion  (Eq. (\ref{drift_diffusion}) ) the two terms on the right-hand side of  Eq. (\ref{drift_energy}) are perfectly balanced and the heating can be expressed directly through
the Lorentz force work $W_D$, which is in that case strictly positive:
\begin{equation}
	Q_{\rm Amb}=D_{\rm Amb}\frac{4\pi}{c^2}\vec{J}_{\perp}^2.\label{qamb_par}
\end{equation}	
However, in the case of the hyperbolic treatment, the term $W_D$ is not strictly positive and Eq. (\ref{qamb_hyp}) is the proper expression to use.

\subsection{Combined treatment of ambipolar diffusion and Hall effect}
\label{sec:amb_hall}
So far our treatment focused only on ambipolar drift and ignored the Hall effect. The Hall effect can be included by substituting in the induction equation
\begin{equation}
   D\vec{v}_D \longrightarrow  D\vec{v}_D -\frac{1}{n_e e}\vec{J}.
\end{equation}
While Hall term can be characterized through the Hall diffusivity
\begin{equation}
    D_{\rm Hall}=\frac{\vert\vec{B}\vert c}{4\pi n_e e},
\end{equation}
the resulting explicit time-step constraints are more severe than the diffusive time-step constraint alone. As described in \citet{Huba:2003:HallMHD}, the Hall term leads to two additional wave modes, the whistler mode and the Hall drift wave in the presence of electron density gradients. Their respective phase speeds are given by
\begin{eqnarray}
    v_{\rm Whistler}&=&k\frac{\vert\vec{B}\vert c}{4\pi n_e e},\\
    v_{\rm Drift}&=&\frac{1}{L_n}\frac{\vert\vec{B}\vert c}{4\pi n_e e},\\
\end{eqnarray}
where the $L_n$ is the density scale length $L_n=(d\ln n_e/d\,x)^{-1}$. For a conservative estimate
we use $L_n>\Delta x$ and $k<2\pi/\Delta x$, leading to an explicit time-step limit of
\begin{equation}
    \Delta t_{\rm Hall}<\frac{\Delta x}{v_{\rm Whistler}+v_{\rm Drift}}<\frac{\Delta x^2}{(\pi+1)D_{\rm Hall}},
\end{equation}
which is a factor of $2-3$ more stringent than the formal (1D) diffusive time-step constraint, but has overall the same functional form of the diffusive time-step limit. Stringent time-step constraints from the Hall term and numerical stability may require special integration procedures as discussed in \citet{Osullivan:2007:STS,Gonzalez:2018:ManchaSTS}. Another possibility is to treat the Hall term similar to the ambipolar drift as part of the hyperbolic time integration. This can be achieved by the following set of equations:
\begin{eqnarray}
    \frac{\partial \vec{v}_D}{\partial t}&=&\frac{1}{\tau_A}\left(\frac{D}{c\varrho_i\nu_{in}}\vec{J}\times\vec{B}-\vec{v}_D\right),\label{ambhallsep1}\\
    \frac{\partial \vec{v}_H}{\partial t}&=&\frac{1}{\tau_H}\left(-\frac{1}{n_e e}\vec{J}-\vec{v}_H\right),\label{ambhallsep2}\\
    \frac{\partial\vec{B}}{\partial t}&=&-\nabla\times\left[\vec{v}\times\vec{B}+(D\vec{v}_D+\vec{v}_H)\times\vec{B}\right],\label{ambhallsep3}
\end{eqnarray}
with
\begin{eqnarray}
    \tau_A&=&\mbox{max}\left(\frac{D}{\nu_{in}},S_A\,f_{\rm N}\frac{D_{\rm Amb}}{C_{\rm MHD}^2}\right),\\
    \tau_H&=&S_H\,f_{\rm N}\frac{D_{\rm Hall}}{C_{\rm MHD}^2}.
\end{eqnarray}
Here we introduced additional control parameters $S_A\ge 1$ and $S_H \ge 1$ that allow us to enhance the averaging time scales if needed for numerical stability. While we did not find a need for values of $S_A$ different from unity, numerical tests presented in Section \ref{ambhall_1d} require values of $S_H$ of up to 5 for numerical stability.

The ratio of the averaging time scales is given by
\begin{equation}
    \frac{\tau_A}{\tau_H}=\frac{S_A}{S_H}\frac{D_{\rm Amb}}{D_{\rm Hall}}=\frac{S_A}{S_H}\frac{D^2 \vert\vec{B}\vert n_e e}{\varrho_i\nu_{in} c} \approx 10^4\frac{S_A}{S_H}\frac{\vert\vec{B}\vert [{\rm G}]}{\nu_{in} [{\rm s}^{-1}]}.\label{tau_ratio}
\end{equation}
In the solar chromosphere we find values of $\nu_{in}$ around $10^4-10^5$ s$^{-1}$(see Figure \ref{fig:1}), which implies that $\tau_H$ and $\tau_A$ are comparable for weaker magnetic field (around 1-10 G), but $\tau_H \ll \tau_A$ in strong field regions. This is the primary reason why ambipolar and Hall drift should be treated in separate relaxation equations as described above. While a combined treatment is possible, it would filter out short-time scale processes resulting from the Hall term since the relaxation time-scale would be dominated by the stronger ambipolar diffusion in many locations.

While the Hall drift is nondissipative, that is not strictly true for the hyperbolic treatment. Since $\vec{v_H}$ can be misaligned with $\vec{J}$, $\vec{v_H}\cdot(\vec{J}\times\vec{B})$ is in general nonzero, although small. $\vec{v_H}$ can have components aligned with $\vec{J}\times\vec{B}$ that mimic ambipolar diffusion and $\vec{v_D}$ can have components aligned with $\vec{J}$ that mimic Hall diffusion. While the latter is (except for the accelerated treatment) physical, the former is an artifact of the numerical treatment introduced here. We will quantify this crosstalk in Section \ref{sec:test2dambhall}.

\section{Note on numerical implementation in the MURaM code}
\label{sec:numerical}
For enhanced stability we integrate the relaxation equations Eq. (\ref{ambhallsep1}) and (\ref{ambhallsep2}) with an additional fourth-order hyperdiffusion added. Since $\tau_A$ and $\tau_H$ vary substantially throughout the computational domain, with very small values in regions where ambipolar diffusion and Hall effect are not time-step limiting, we use a point-implicit time integration for numerical stability, i.e. we solve numerically
the following equations:
\begin{eqnarray}
    \vec{v}_D(t+\Delta t)&=&\frac{\vec{v}_D(t)-0.02\;\Delta^2 \vec{v}_D(t)}{1+\Delta t/\tau_A}+\frac{\Delta t/\tau_A}{1+\Delta t/\tau_A}\left(\frac{D}{c\varrho_i\nu_{in}}\vec{J}\times\vec{B}\right)(t),\\
    \vec{v}_H(t+\Delta t)&=&\frac{\vec{v}_H(t)-0.02\;\Delta^2 \vec{v}_H(t)}{1+\Delta t/\tau_H}+\frac{\Delta t/\tau_H}{1+\Delta t/\tau_H}
    \left(-\frac{1}{n_e e}\vec{J}\right)(t),
\end{eqnarray}
where $\Delta^2$ denotes the fourth derivative in grid space, i.e. $\Delta^2f=f(i-2)-4 f(i-1)+6 f(i)-4 f(i+1)+f(i+2)+f(k-2)+[\ldots]$. 

In addition, we modify the characteristic velocity that is used to compute numerical diffusivities for the remainder of the MHD system \citep{rempel2014:ssd} by adding both $v_D$ and $v_H$. The time integration for the ambipolar drift is implemented as part of the four-step MURaM time-integration scheme and spatial derivatives are computed with the fourth-order centered finite difference operator as described in \citet{Voegler:2005}. Note that the time-integration scheme does allow for Courant–Friedrichs–Lewy (CFL) numbers of up to $2$.

\section{Tabulated collision frequencies}
\label{sec:tabs}
To calculate the collisional frequencies we follow the method described by \citet{Nobrega:2020:BifrostSTS}. The MURaM code uses a pretabulated LTE equation of state (EoS). The EoS is calculated using the freeEoS \citep{Irwin:2012:FreeEoS} package, which includes 20 of the most abundant elements in the solar atmosphere. Earlier MURaM models \citep{Rempel:2017:corona} used a combination of OPAL \citep{rogers:1996:OPAL} and the \citet{gustafsson:1975:EOS} EoS, mostly since OPAL does not extend into the higher atmosphere and \citet{gustafsson:1975:EOS} showed differences in the convection zone. The freeEoS package \citep{Irwin:2012:FreeEoS} provides tables that are very similar to OPAL in the convection zone and extend into the corona, while having overall smoother tables with fewer numerical artifacts. We use the abundances of \citet{gustafsson:1975:EOS} in order to keep backwards compatibility with existing simulations, but have also tables with newer abundances available. A detailed comparison of the impact different abundances have on the chromosphere and corona is a worthwhile investigation as, for example, \citet{Asplund:2009} gives 5\% more hydrogen per gram and a lower helium abundance. However, this will effect many details, from opacities, to the cooling in the strong lines of the chromosphere, in nonequilibrium hydrogen and also ion-neutral effects. It is beyond the scope of this paper to investigate the effect of abundances in detail. The molecules $H_2$, and $H_2^+$ are included. The ion-density is then calculated
\begin{equation}
\varrho_i = \sum_{a=1}^{20}\sum_{i=1}^{9} n_{a,i} m_{a}.
\end{equation}

A range of approximations exist to calculate the ion-neutral cross sections and collisional rates. A comparison of the different approaches and the resulting differences in a 2D RMHD simulation was investigated by \citet{martinezsykora:2012:PI_1}. We follow the method described by \citet{Nobrega:2020:BifrostSTS}, which includes modern atomic data.

The collisions between neutrals and ions are calculated as
\begin{equation}
    \nu^*_{ni} = \frac{m_{ni}}{m_n}n_i\sigma_{ni}\left(\frac{8 k_B T}{\pi m_{ni}}\right)^{1/2},
\end{equation}
where $k_B$ is Boltzmann's constant and $m_{ni}=\frac{m_n m_i}{m_n + m_i}$ is the reduced mass. We consider collisions between neutral hydrogen, neutral helium, and molecular H$_2$ with ions of all 20 elements in the equation of state. For p-H~{\sc i}, p-He~{\sc i}, p-H$_2$, e-H~{\sc i}, e-He~{\sc i}, e-H$_2$, and He~{\sc ii}-He~{\sc i} collisions we use the temperature dependent cross sections described in \citet{Nobrega:2020:BifrostSTS}, and references within. For collisions between H~{\sc i}, He~{\sc i} and metal ions $(m)$ we follow the same assumption of \citet{vranjes:2008:PI}. The cross section between a neutral and a metal is given by $\sigma_{nm} = \frac{m_m}{m_n} \sigma_{np}$, where $\sigma_{np}$ is the cross section between neutral hydrogen, or helium, and protons. 

\begin{figure*}
    \centering
    \resizebox{0.95\hsize}{!}{\includegraphics{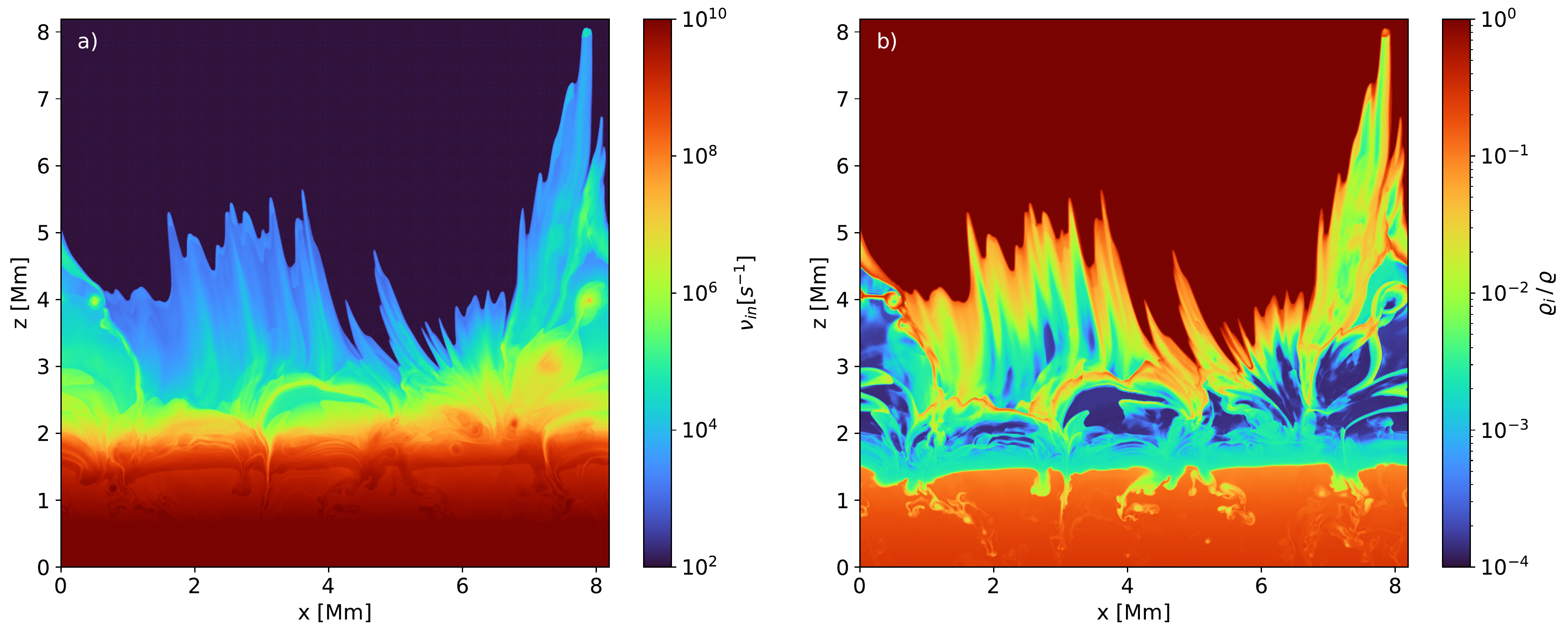}}
    \caption{Values of a) $\nu_{in}$ and b) $\varrho_i/\varrho$ for a snapshot from the 2D test setup that will be discussed further in Section \ref{sec:test_2D}.}
    \label{fig:1}
\end{figure*}

\begin{figure*}
    \centering
    \resizebox{0.95\hsize}{!}{\includegraphics{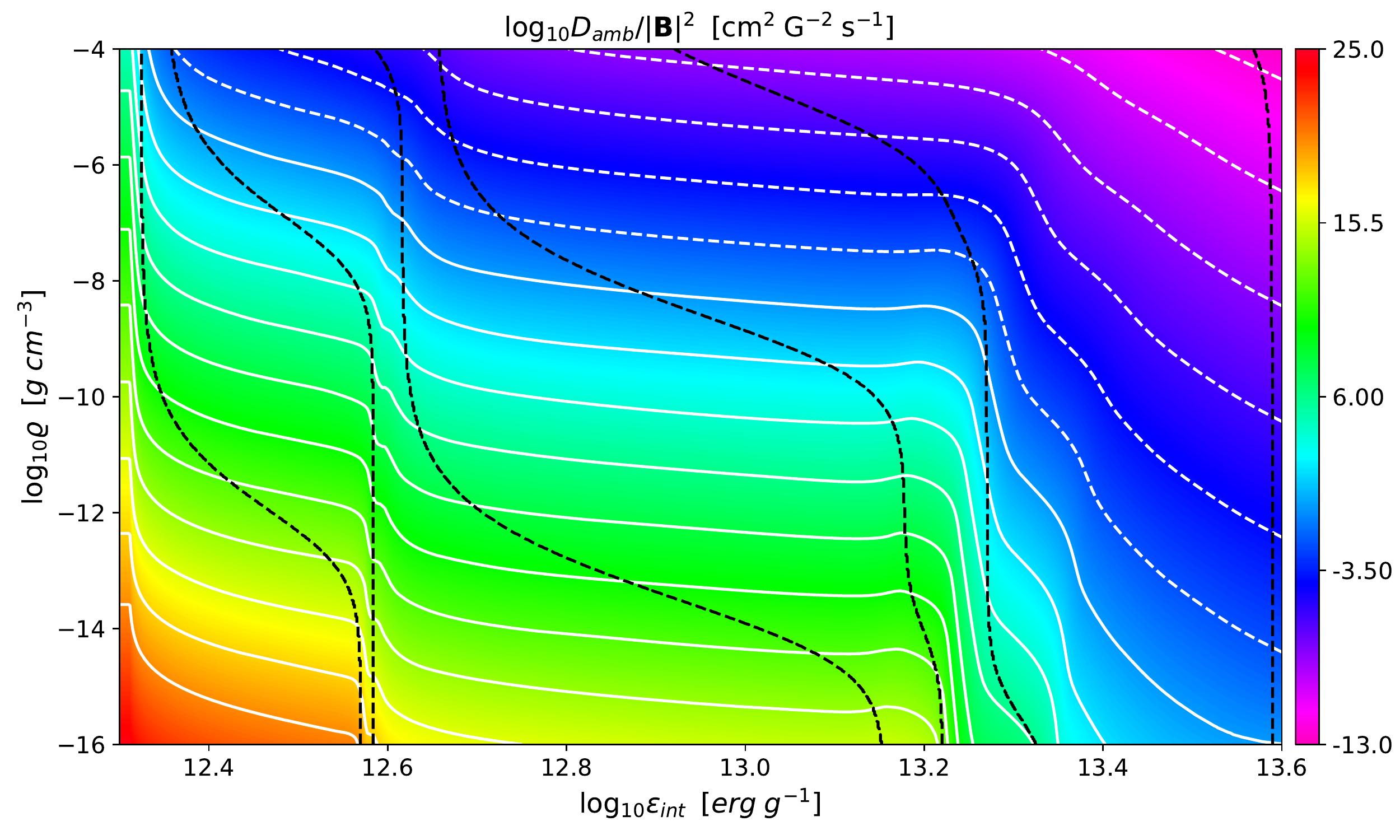}}
    \caption{Values of $D_{\rm Amb}$ as a function of density $\rho$ and internal energy $\epsilon_{int}$ for the EoS table. The internal energy has been shifted to match the figure of \citep{Nobrega:2020:BifrostSTS}. Black dashed lines show temperature contours of $1.8\times 10^3~\mathrm{K}$, $3\times 10^3~\mathrm{K}$, $6\times 10^3\;\mathrm{K}~\mathrm{K}$, $1\times 10^4~\mathrm{K}$, $2\times 10^4~\mathrm{K}$, and $1\times 10^5~\mathrm{K}$. White lines show contours of $D_{\rm Amb}$ separated by two orders of magnitude.}
    \label{fig:2}
\end{figure*}

The average neutral-ion collision frequency is
\begin{equation}
\rho_n \nu_{ni} = \varrho_{\mathrm{H}\;\textsc{i}} \nu^*_{\mathrm{H}\;\textsc{i},\mathrm{e}} + \varrho_{\mathrm{He}\;\textsc{i}} \nu^*_{\mathrm{He}\;\textsc{i},\mathrm{e}} + \varrho_{\mathrm{H}_2}\left(\nu^*_{\mathrm{H}_2,p} + \nu^*_{\mathrm{H}_2,\mathrm{e}}\right) + \sum_{i=0}^{20} \left(\varrho_{H\;\textsc{i}} \nu^*_{\mathrm{H}\;\textsc{i},i} +\varrho_{\mathrm{He}\;\textsc{i}} \nu^*_{\mathrm{He}\;\textsc{i},i} \right).
\end{equation}
The average ion-neutral collision frequency $\nu_{in}$ is then calculated from conservation of momentum, $\varrho_i \nu_{in} = \varrho_n \nu_{ni}$.  The values of $\nu_{in}$ and $\rho_i$ calculated from a 2D simulation of the solar atmosphere are shown in Figure \ref{fig:1}. The collisional rates in the cold intershock regions are approximately two orders of magnitude higher than using the cross sections of \citet{Osterbrock:1961} and \citep{dePontieu:Haerendel:1998}, as used previously in \citet{Cheung:Cameron:2012:muram_amb_hall}. These differences arise from the approximation used for the neutral-metal cross sections, which are significantly larger than those used before. Finally, $D_{\rm Amb}$ is calculated from Eqn. \ref{ambdiff}. Figure \ref{fig:2} shows the tabulated values of $D_{\rm Amb}$, reproducing the figure in Section 4.1 of \citet{Nobrega:2020:BifrostSTS}.

\section{Numerical tests}
\label{sec:tests}
\subsection{1D Idealized Ambipolar diffusion test}
\label{sec:1D_Amb}

\begin{figure*}
    \centering
    \resizebox{0.95\hsize}{!}{\includegraphics{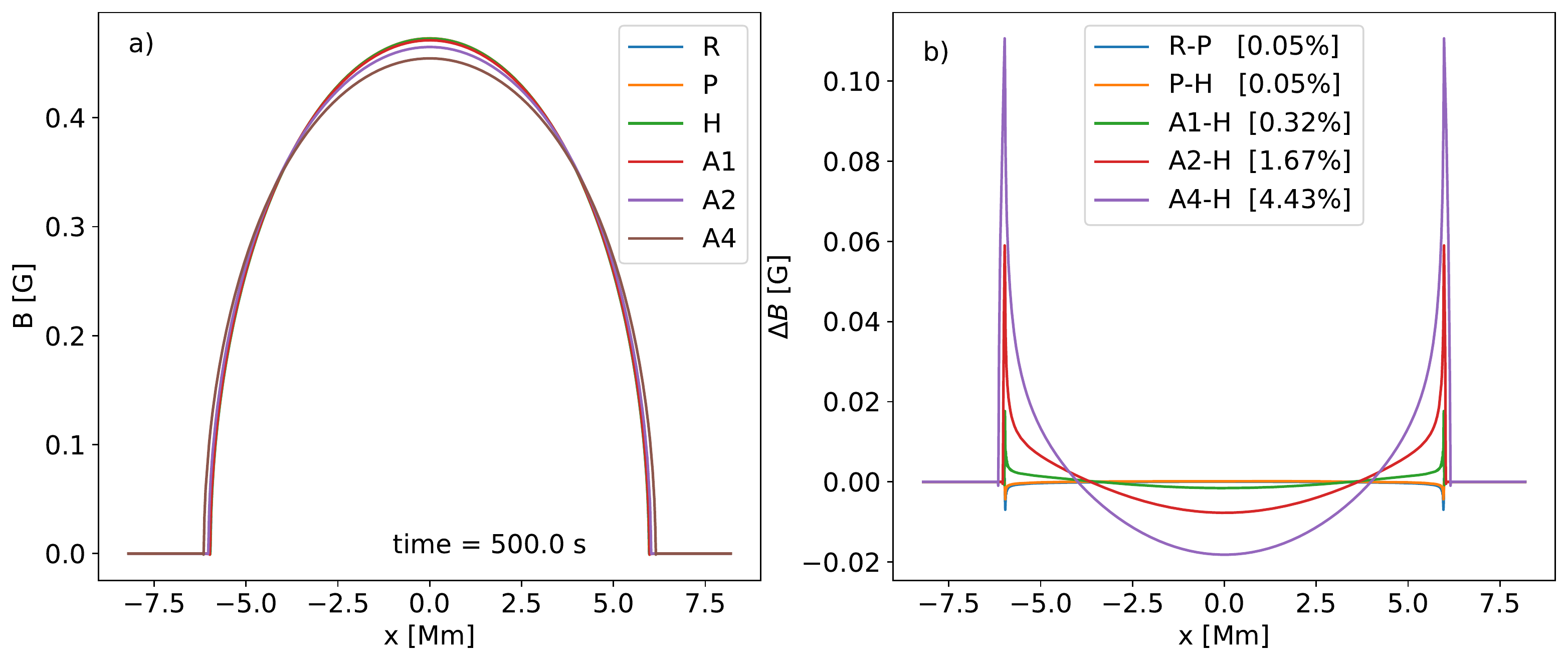}}
    \caption{1D diffusion test problem after 500 seconds of evolution. Panel (a): analytical asymptotic self-similar solution (R), explicit parabolic treatment (P), hyperbolic treatment (H), and accelerated hyperbolic treatments (A) with time-step limits of $0.1$, $0.2$, and $0.4\, \Delta t_{\rm Drift}$. Panel (b): differences between the solutions as indicated. Values in brackets show the relative RMS errors. The solution (H) was integrated 16 times faster as (P), the solutions
	(A1), (A2), and(A4) were integrated 73, 144, and 250 times faster than (P). }
    \label{fig:3}
\end{figure*}
We first test the hyperbolic and accelerated hyperbolic treatments on a 1D ambipolar diffusion setup. We choose a 16.384 Mm wide domain with 8 km grid spacing and assume a fixed value of $\varrho_i=10^{-17}$ g cm$^{-3}$ and $\nu_{in}=10$ s$^{-1}$, leading to a constant ambipolar diffusivity of $\eta_{\rm Amb}=10^{16}/4\pi$ cm$^2$ s$^{-1}$ G$^{-2}$ (we assume here $D=1$). The parabolic solution follows from the equation
\begin{equation}
	\frac{\partial B_y}{\partial t}=\eta_{\rm Amb}\frac{\partial}{\partial x}\left(B_y^2\frac{\partial B_y}{\partial x}\right).
\end{equation}
We initialize the problem with a Gaussian profile:
\begin{equation}
	B_y(t=0)= B_0 \exp{\left[-x^2/\sigma^2\right]},
\end{equation}
with $B_0=5$~G and $\sigma=0.5$~Mm. The asymptotic self-similar solution is given by \citep{Pattle:1959}
\begin{eqnarray}
	R(t)&=&\left[\left(\frac{2\Phi}{\pi}\right)^2\;4\eta_{\rm Amb}\;t \right]^{\frac{1}{4}},\\
	B_y(t)&=&\frac{2\Phi}{\pi}\frac{1}{R(t)}\sqrt{1-\frac{x^2}{R(t)^2}}\;.
\end{eqnarray}
Here $\Phi=\int B_y(x) dx$ denotes the constant 1D "magnetic flux" of the solution. The peak field strength of the self-similar solution is given by $\frac{2\Phi}{\pi}\frac{1}{R(t)}$. 

For the parabolic treatment only the product $\varrho_i \nu_{in}$ determines the ambipolar diffusivity, but the individual choices for  $\varrho_i$ and $\nu_{in}$ do affect the hyperbolic treatment. For example, the combination of the very small $\varrho_i$  and large $\nu_{in}$ is more in favor of a parabolic treatment once the resulting large values of $\vert\vec{B}\vert/\sqrt{4\pi\varrho_i}$ become time-step limiting. The opposite situation is, however, found in solar applications (see Section \ref{sec:test_2D} for further detail). We aim at capturing that through our choice. The initial ambipolar diffusivity of the setup is $2\times 10^{16}$cm$^2$ s$^{-1}$, the initial value of $\vert\vec{B}\vert/\sqrt{4\pi\varrho_i}=4460$ km s$^{-1}$, leading initially to a diffusive time-step limit of $1.6\times 10^{-5}$ s and a hyperbolic time-step limit of $1.8\times 10^{-3}$ s, i.e. a speedup of more than 2 orders of magnitude. Over the 500 seconds of simulated evolution the speedup was 15 times. How much additional speedup the accelerated hyperbolic treatment provides depends on the ratio of $\vert\vec{v}_D\vert$ to $\vert\vec{B}\vert/\sqrt{4\pi\varrho_i}$. Since the self-similar solution has very steep gradients, $\vert\vec{v}_D\vert$ did reach initially values of up to $1300$ km s$^{-1}$, but dropped toward the end to values as low as $3$ km s$^{-1}$, while $\vert\vec{B}\vert/\sqrt{4\pi\varrho_i}$ dropped to $420$ km s$^{-1}$. 

For the accelerated hyperbolic treatment we consider three cases with time steps relative to ${\Delta t}_{\rm Drift}=\Delta x/ \vert\vec{v}_D\vert$ (since the ambipolar diffusion problem does not have an MHD time step we can use as reference) of $0.1$, $0.2$ and $0.4$, leading to overall speedups of $73$, $144$, and $250$ times, respectively. Figure \ref{fig:3} compares all these setups at a time of $500$ seconds. Panel (a) shows the profile of $B_y$, while panel (b) shows the differences and relative RMS errors as indicated. The parabolic solution (P) is within  $0.05\%$ of the asymptotic reference solution (R). They are not expected to be identical since (1) we did not initialize  the simulation with a self-similar profile and (2) the numerical treatment includes additional numerical diffusivities that affect in particular the regions with steep gradients near the edge of the self-similar profile. The hyperbolic solution (H) is within $0.05\%$ of the parabolic one, which essentially emphasizes that the hyperbolic treatment behaves overall as the parabolic one, while allowing for more than an order of magnitude speedup (for the particular setup considered here). Note that the hyperbolic solution is more correct as it does consider the time derivative of the drift velocity that is neglected in the parabolic treatment. The accelerated hyperbolic solutions (A1, A2, A4) differ by $0.3-4.4\%$ from the hyperbolic (H) reference.  While the accelerated treatment allows for a substantial increase in speed by up to 250 times in the case of (A4), there is a trade-off between speed and accuracy. All the accelerated hyperbolic solutions do show a stronger spread as compared to (H), (P) and (R), indicating that the accelerated treatment does lead to a small overestimation of the ambipolar drift. 

The 1D test problem does only consider the ambipolar drift in separation. In a realistic MHD application additional time-step constraints arise from the MHD system and in many cases the amplitude of  $\vert\vec{v}_D\vert$ remains small compared to the characteristic speeds of the MHD system, which limits naturally the accelerated hyperbolic treatment to $\Delta t < 0.1\,\Delta x/\vert\vec{v}_D\vert$, i.e. errors arising from the accelerated hyperbolic treatment
do remain small.

\begin{figure}
        \centering
        \resizebox{0.85\hsize}{!}{\includegraphics{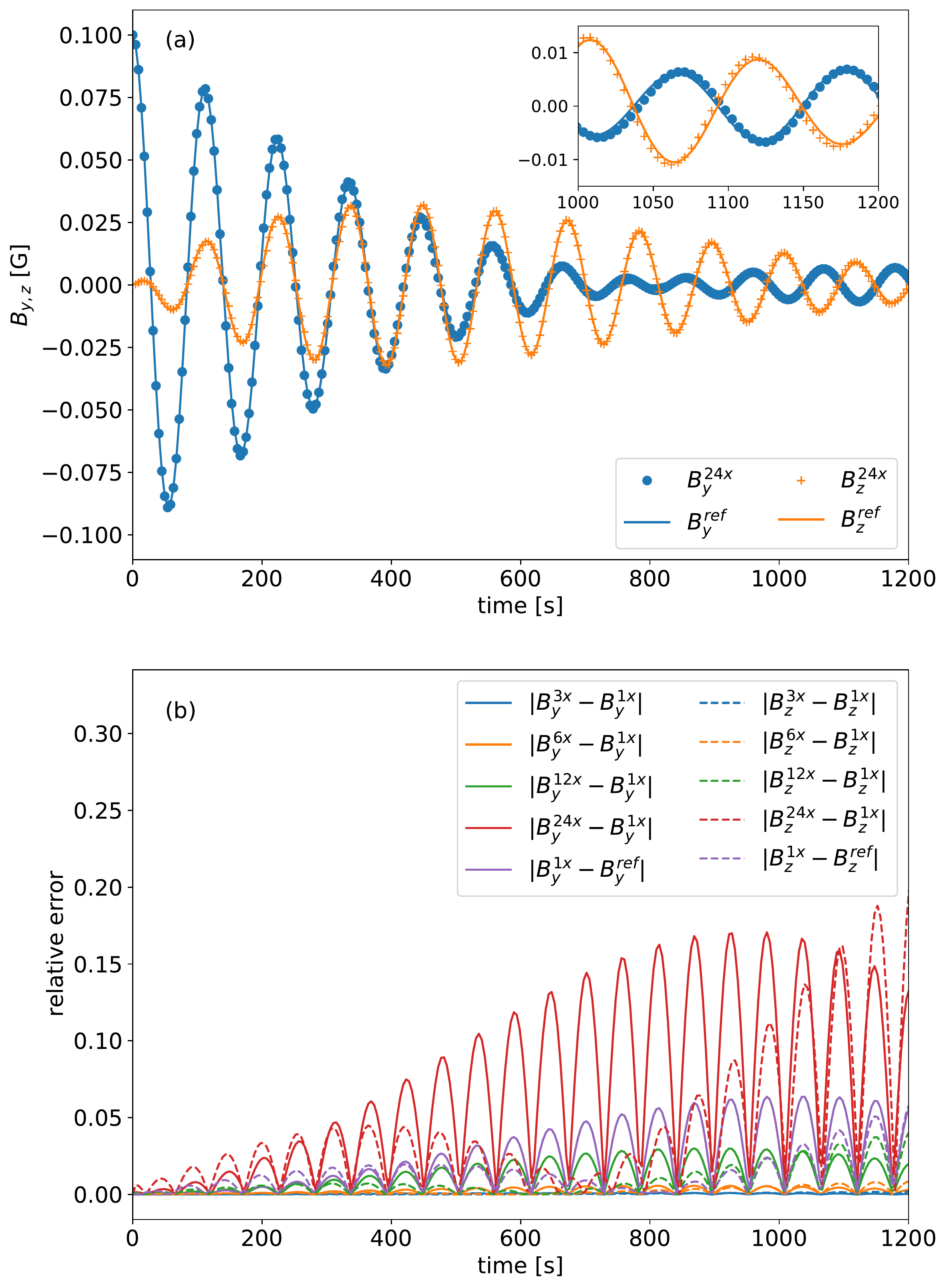}}
        \caption{1D damped standing wave solution with combined action of ambipolar diffusion and Hall effect. We show the quantities $B_y(x=0,t)$ and $B_z(x=0,t)$. (a) Numerical solution from the $24\times$ accelerated hyperbolic treatment in comparison to the approximate analytical solution. The dominant difference is a phase shift. (b) Error of various accelerated hyperbolic solutions relative to the nonaccelerated solution. The error is relative to the decaying envelope of the solution $b_1\exp(-\gamma t)$. In addition we show the difference of the nonaccelerated numerical and approximate analytical solutions.}
        \label{fig:4}
\end{figure}

\subsection{Influence of Hall effect and ambipolar diffusion on 1D wave propagation}
\label{ambhall_1d}
The next test problem is a generalization of the Hall test problem from \citet{Cheung:Cameron:2012:muram_amb_hall,Gonzalez:2018:ManchaSTS}. See also \citet{Cally:Khomenko:2015} for a more accurate interpretation of the underlying wave dynamics. We include in addition to the Hall effect also ambipolar diffusion. We solve
\begin{eqnarray}
    \varrho_0\frac{\partial \vec{v}}{\partial t}&=&\frac{1}{4\pi}(\nabla\times\vec{B})\times\vec{B},\\
    \frac{\partial \vec{B}}{\partial t}&=&\nabla\times\left[\vec{v}\times\vec{B}-H(\nabla\times\vec{B})\times\vec{B}+A\left(\nabla\times\vec{B})\times\vec{B}\right)\times\vec{B}\right].
\end{eqnarray}
Using an ansatz and setup similar to \citet{Cheung:Cameron:2012:muram_amb_hall} (i.e., only $x$-dependence, but also considering ambipolar diffusion and the resulting damping of the solution), we find a damped standing wave solution of the form (we use $b_1=10^{-3} b_0$ in order to remain in the linear regime)
\begin{eqnarray}
    B_x &=& b_0,\\
    B_y &=& b_1\cos(\sigma t)\cos(k x)\cos(\omega t)\exp(-\gamma t)\label{Byexact},\\
    B_z &=& b_1\sin(\sigma t)\cos(k x)\cos(\omega t)\exp(-\gamma t),\label{Bzexact}
\end{eqnarray}
where
\begin{eqnarray}
    \gamma &=& \frac{1}{2}\eta_A k^2,\\
    \sigma &=& \frac{1}{2}\eta_H k^2,\\
    \omega &=& \sqrt{(v_A k)^2+\sigma^2-\gamma^2},\\
\end{eqnarray}
and $\eta_A=A b_0^2$, $\eta_H=H b_0$, and $v_A=b_0^2/4\pi\varrho_0$. Note that these expressions are approximate, we neglected cross terms that couple $\sigma$ and $\gamma$. If either $\sigma$ or $\gamma$ are zero, these expressions are exact. Since we consider here a situation with $\sigma\,,\gamma \ll \omega$, this solution is sufficiently accurate as reference for the numerical test.

As in \citet{Cheung:Cameron:2012:muram_amb_hall} we use a domain extending $100$~km in $x$, and we choose $k=2\pi\times 10^{-2}$ km, $b_0 = 100$~G, and $\varrho_0=10^{-7}$ g cm$^{-3}$. We pick $\eta_H=\eta_A=10^{10}$ cm$^2$ s$^{-1}$. These choices lead to $\omega=0.056$ s$^{-1}$ and $\gamma=\sigma=0.002$ s$^{-1}$. The numerical solution is initialized with $\vec{B}=[b_0,b_1\cos(kx),0]$ and $\vec{v}=[0,\gamma,-\sigma] b_1/(k b_0) \sin(kx)$.
We pick a grid spacing of $\Delta x=100$~m, which leads to explicit diffusive time-step limits of $0.5\,\Delta x^2/\eta_A = 0.5\,\Delta x^2/\eta_H = 0.005$ s for ambipolar and Hall diffusion (the latter could be a factor of 2-3 more stringent if we adopt a conservative estimate for the whistler mode). With a safety factor of $0.75$ an explicit integration would be limited to time steps of $0.00375$ s. We use the hyperbolic treatment to speed up the integration by a factors of $3$, $6$, $12$ and $24$ over this baseline, allowing for a time step as large as $0.09$ s. Accelerations much larger than $24\times$ are not possible in this setup, as we reached the MHD time step . While we use $S_A=1$, stability of the hyperbolic treatment of the Hall effect requires a value of $S_H\approx\sqrt{\mbox{speedup}}$, i.e. values of $1.7$, $2.5$, $3.5$, and $4.9$. These settings are specific for this problem and depend also critically on the value of the ambipolar diffusivity. A lower value of $\eta_A$ would require larger values of $S_H$ for stability. Too small values of $S_H$ lead to an initially slow growing instability manifest in the shortest wavelength resolvable on the numerical grid, which eventually destroys the coherence of the solution. However, this instability does not lead to exponential error growth and the solution remains bounded. In the case
of the 2D solar test setups presented in Section \ref{sec:test_2D} we do not find instabilities using $S_H=1$. This is likely due to a moderate speedup of the Hall treatment by about a factor of $7$ in combination with a significantly larger ambipolar diffusivity.

$24\times$ accelerated solutions for $B_y(x=0,t)$ and $B_z(x=0,t)$ are given in Figure \ref{fig:4} (a) together with the approximate analytical solution from Eqs. (\ref{Byexact}) and (\ref{Bzexact}). The insert panel shows the solution for the times from $1000$ to $1200$ s. The
dominant error is a phase shift, while the amplitude of the solution is well captured. The phase shift is comparable to $\tau_H\approx 1.7$ s, with the solution for $B_y$ trailing and the solution for $B_z$ leading the reference solution. In panel (b) we show the errors of the $3\times$, $6\times$, $12\times$, and $24\times$ accelerated solutions relative to the nonaccelerated ($1\times$) numerical solution in order to separate errors resulting from the hyperbolic treatment from other numerical effects as well as the approximate nature of analytical reference. The difference between the nonaccelerated and approximate analytical solution is on the order of $5\%$, which is larger than the differences found in the $3\times$, $6\times$, and $12\times$ accelerated solutions relative to the $1\times$ baseline.

\subsection{2D solar setup}
\label{sec:test_2D}

\subsubsection{Hyperbolic ambipolar drift}
\begin{figure*}
        \centering
        \resizebox{0.95\hsize}{!}{\includegraphics{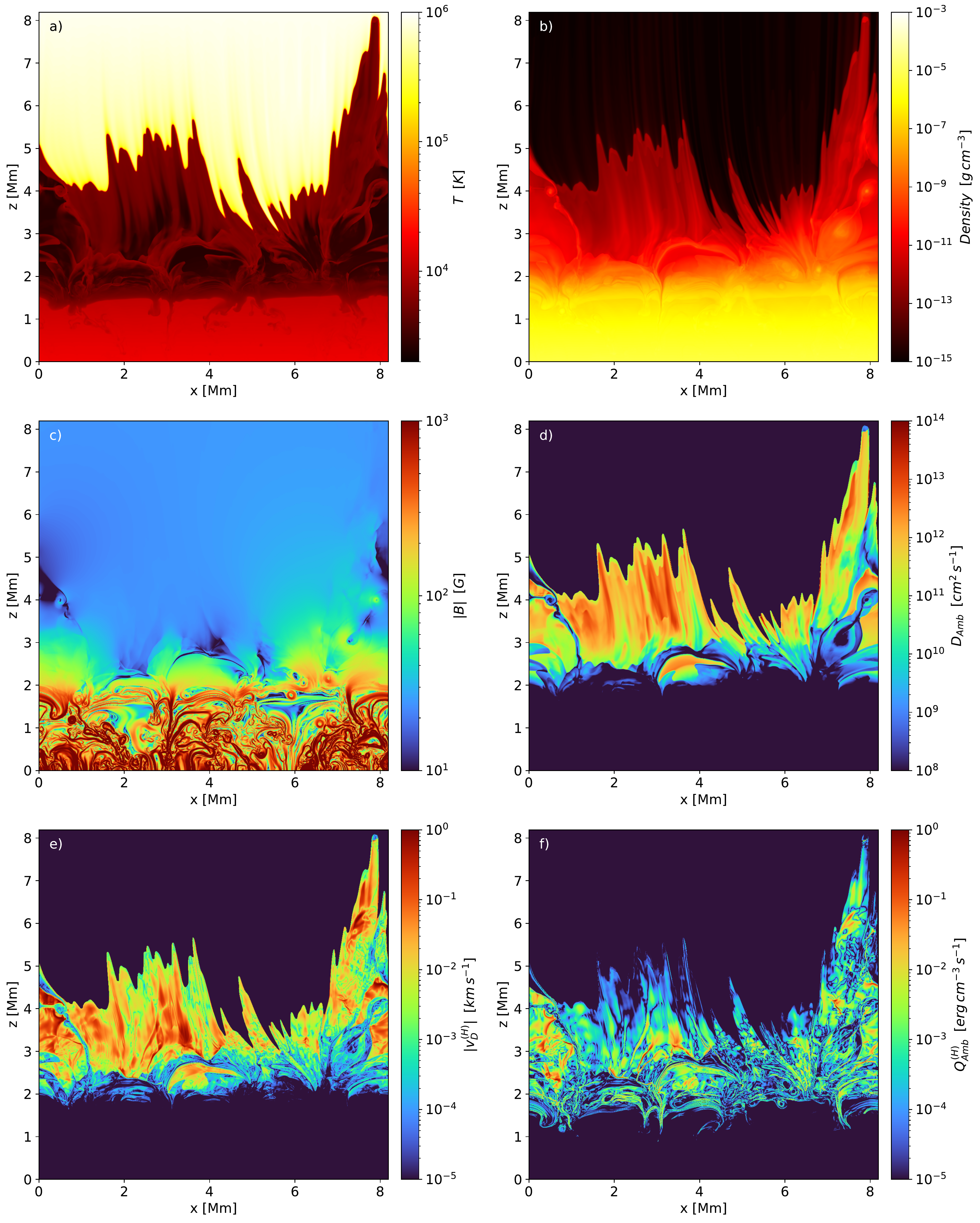}}
        \caption{Snapshot from the test setup. Presented are a) temperature, b) density, c) magnetic field strength, d) ambipolar diffusivity, e) ambipolar drift velocity, and f) ambipolar heating. The latter two were computed using the hyperbolic treatment Eqs. (\ref{drift3}, \ref{qamb_hyp}).}
        \label{fig:5}
\end{figure*}

In order to test our hyperbolic implementation of ambipolar drift, we use two-dimensional simulation setups in a $8.192\times8.192$ Mm domain with 8 km grid spacing. The simulation domain starts 1.5 Mm beneath the photosphere and reaches about 6.5 Mm above the photosphere. At the top boundary we impose a temperature of 1 million K in order to create a transition region above the chromospheric part of the domain. To this end we include besides MHD and gray radiative transfer also Spitzer heat conduction and CHIANTI based optically thin radiative losses as described in \citet{Rempel:2017:corona}. The simulations are integrated with $S_A=1$, $S_H=1$ and a CFL number of 1.5. Initially we imposed in our setup a $25$ G uniform vertical magnetic field and evolved the simulation until the field was concentrated in the convective downflow region. Figure \ref{fig:5} shows a snapshot from the setup, with panel (a) showing temperature and panel (b) showing the corresponding mass density. Panel (c) shows the magnetic field strength. The resulting ambipolar diffusion constant (see Eq. (\ref{ambdiff}))  is presented in panel (d). Panels (e) and (f) show the ambipolar drift velocity and ambipolar heating, respectively. This simulation was integrated using the hyperbolic treatment Eq. (\ref{drift3}).

\begin{figure*}
        \centering
        \resizebox{0.95\hsize}{!}{\includegraphics{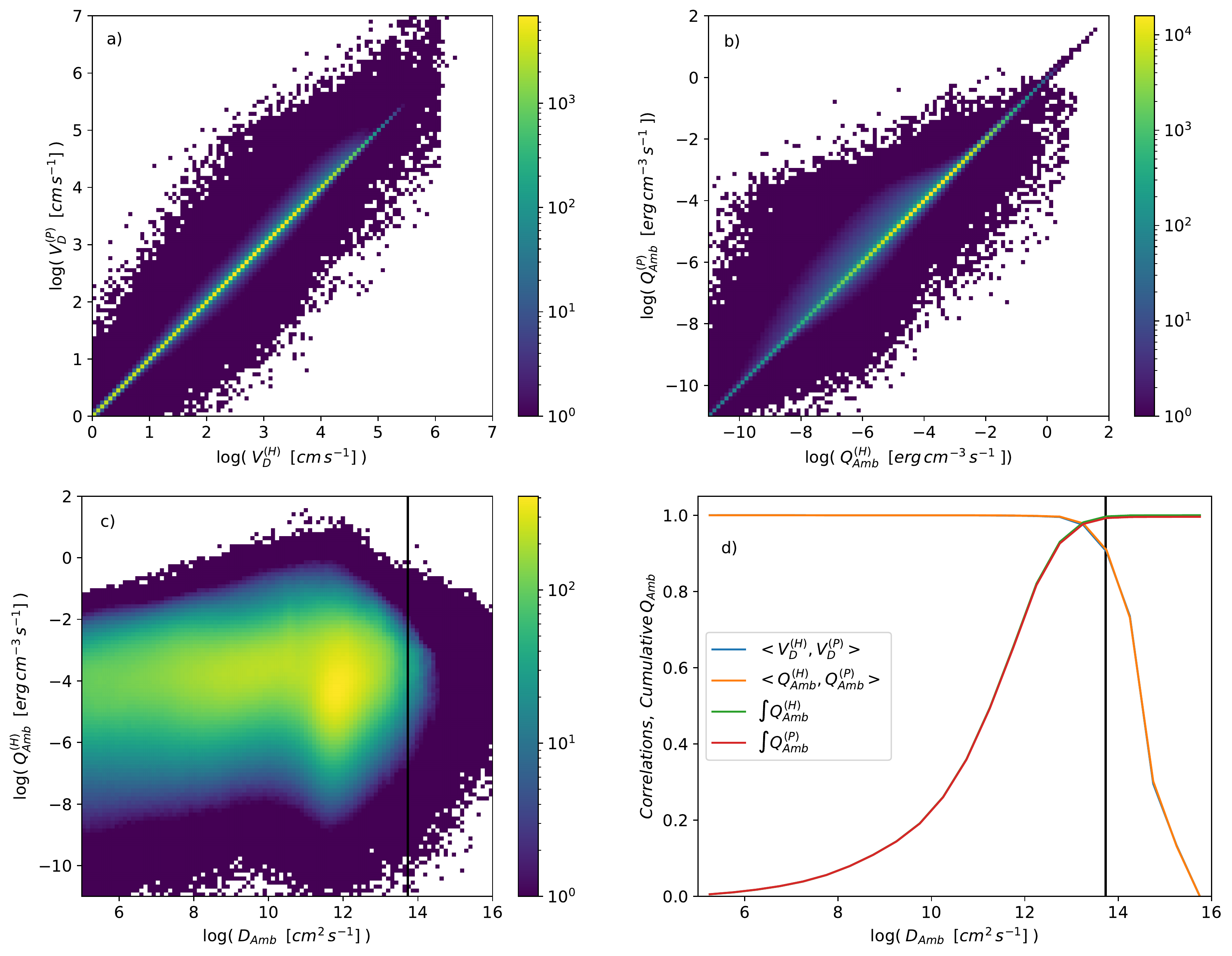}}
    \caption{Statistical comparison between hyperbolic and parabolic treatment. Panels (a) and (b): 2D histogram showing the $V_D^{(H)}$ - $V_D^{(P)}$ and  $Q_{\rm Amb}^{(H)}$ - $Q_{\rm Amb}^{(P)}$ correlations. Overall these these quantities remain correlated at the $99.9\%$ level. Panel (c) 2D histogram of the $D_{\rm Amb}$ - $Q_{\rm Amb}^{(H)}$ correlation. The most significant contributions to ambipolar heating come from regions with moderate values of diffusivity. Panel (d): $V_D^{(H)}$ - $V_D^{(P)}$ and  $Q_{\rm Amb}^{(H)}$ - $Q_{\rm Amb}^{(P)}$ correlation as afunction of $D_{\rm Amb}$ (blue ond orange) and cumulative ambipolar heating computed through $Q_{\rm Amb}^{(H)}$ (green) and $Q_{\rm Amb}^{(P)}$ (red). The vertical black line in panels c) and d) indicates the $D_{\rm Amb}$ value at which the hyperbolic treatment takes over.}
        \label{fig:6}
\end{figure*}

\begin{figure*}
        \centering
        \resizebox{0.95\hsize}{!}{\includegraphics{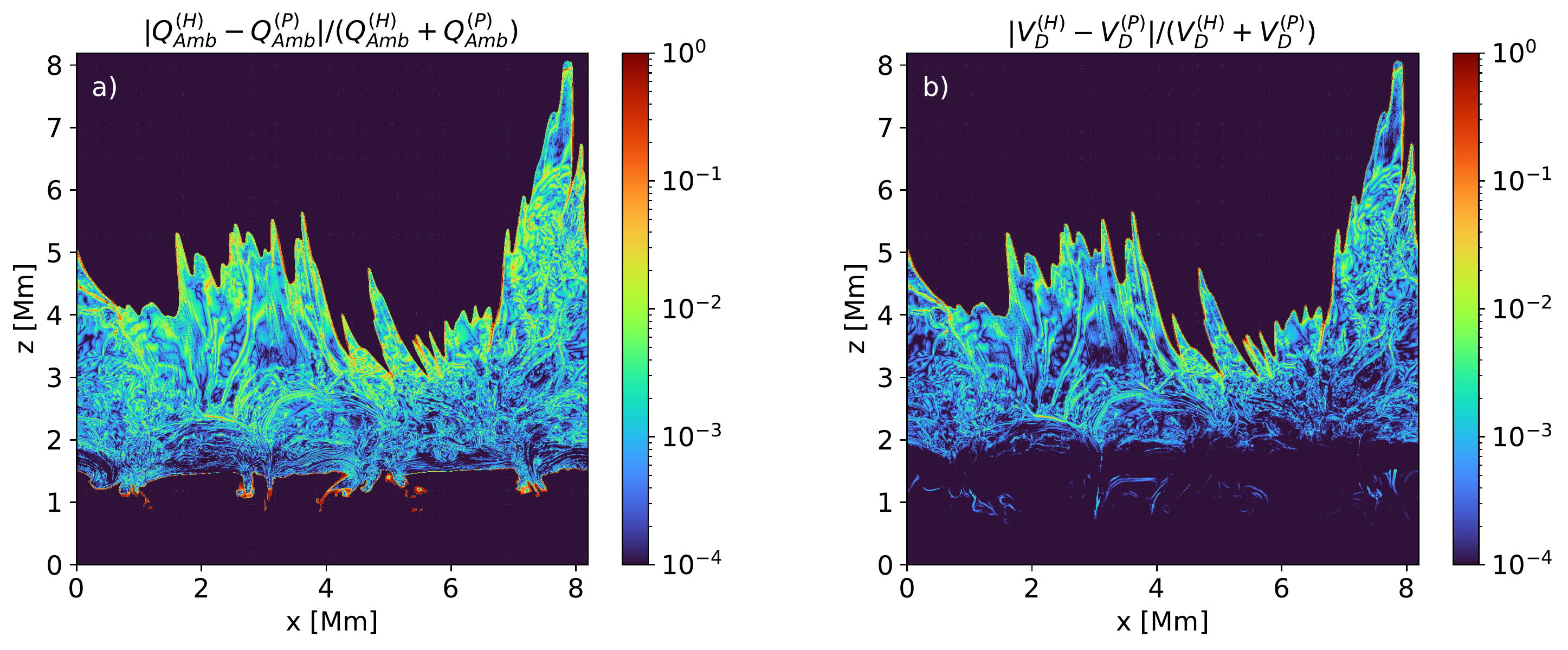}}
        \caption{Spatial distribution of relative differences between hyperbolic and parabolic quantities. (a) Difference between $Q_{\rm Amb}^{(H)}$ and $Q_{\rm Amb}^{(P)}$; b) difference between  $V_D^{(H)}$ and $V_D^{(P)}$. These quantities are shown for the same snapshot as presented in Figure \ref{fig:5}. Relative differences are generally larger in the upper chromosphere, but remain mostly at the percent level.}
        \label{fig:7}
\end{figure*}        

We compare the parabolic and hyperbolic treatment by analyzing a simulation run that covers about 25 minutes of temporal evolution. We evolved the simulation using the hyperbolic treatment and compare the resulting
ambipolar drift and heating with those expected from the diffusive treatment. Peak values of the ambipolar diffusivity are mostly in the range of $10^{14} - 10^{16}$ cm$^2$ s$^{-1}$, with extreme values reaching $10^{17}$ cm$^2$ s$^{-1}$. This leads to explicit diffusive time-step constraints mostly in the range of $10^{-3}$ down to $10^{-6}$ seconds, with an average diffusive time-step constraint of about $5\times 10^{-5}$ over the duration of the simulation. Typical values of $C_{\rm Amb}^{\rm hyp}=\vert\vec{B}\vert\sqrt{D}/\sqrt{4\pi\varrho_i}$ are around $8000$ km s$^{-1}$, the resulting time step of hyperbolic treatment was on average $0.0015$ s. Over the duration of the simulation the hyperbolic treatment provides a speedup of more than $10$ times over the parabolic treatment and even more in a few extreme cases. Since the value of the Alfv{\'e}n velocity in the coronal part of the simulation domain also reaches values as high as $6000$ km s$^{-1}$, the hyperbolic treatment naturally allows us to evolve the system close to the native MHD time step. The benefit of the hyperbolic treatment would be even larger for simulations with stronger field and higher resolution, as well as larger-scale 3D simulations, in which extreme values of the ambipolar diffusivity would be realized in almost every time step.

In Figure \ref{fig:6} we compare the hyperbolic and parabolic ambipolar treatment. Here $V_D^{(P)}$ and $Q_{\rm Amb}^{(P)}$ denote the ambipolar drift velocity amplitude and heating computed using the parabolic treatment, 
Eq. (\ref{drift_diffusion}) and Eq. (\ref{qamb_par}), whereas $V_D^{(H)}$ and $Q_{\rm Amb}^{(H)}$ denote the respective quantities that follow from Eq. (\ref{drift3}) and Eq. (\ref{qamb_hyp}). Note that we did evolve the simulation using the hyperbolic treatment, but computed the parabolic quantities from the simulation snapshots for comparison. In Figure \ref{fig:6}  panels (a) and (b) we show 2D histograms of the $V_D^{(H)}$ - $V_D^{(P)}$ and  $Q_{\rm Amb}^{(H)}$ - $Q_{\rm Amb}^{(P)}$ scatter. Both quantities are strongly correlated at a level above $99.9\%$. Panel (c) shows the distribution of ambipolar heating with ambipolar diffusivity. The largest contribution comes from regions that do have diffusivities in the range from $10^{11}-10^{13} {\rm cm}^2 {\rm s}^{-1}$, while peak values of $D_{\rm Amb}$ can exceed $10^{16} {\rm cm}^2 {\rm s}^{-1}$. Panel (d) shows the correlation coefficients $V_D^{(H)}$ - $V_D^{(P)}$ (blue) and $Q_{Amb}^{(H)}$ - $Q_{\rm Amb}^{(P)}$ (orange) as a function of ambipolar diffusivity. The green and red curves show the cumulative values of $Q_{\rm Amb}^{(H)}$ and $Q_{\rm Amb}^{(P)}$ up to the respective $D_{\rm Amb}$ values. In panels (c) and (d) the vertical black line indicates the threshold where the hyperbolic treatment essentially takes over, i.e. the $D_{\rm Amb}$ value for which the diffusive time step constraint is identical to the hyperbolic one (about $5\times 10^{13} {\rm cm}^2 {\rm s}^{-1}$). As expected we find that the hyperbolic treatment is close to identical to the parabolic treatment up to this threshold and the correlations $V_D^{(H)}$ - $V_D^{(P)}$ and $Q_{\rm Amb}^{(H)}$ - $Q_{\rm Amb}^{(P)}$ drop significantly above. Most of the cumulative ambipolar heating is accounted for already at the threshold. The lower correlations above the threshold do not point toward inaccuracies in the hyperbolic treatment, to the contrary, they indicate that the partial time derivative of the drift velocity does make a significant contribution and cannot be neglected as done in the parabolic treatment. The spatial distribution of the relative differences between $Q_{\rm Amb}^{(H)}$ and $Q_{\rm Amb}^{(P)}$ as well as $V_D^{(H)}$ and $V_D^{(P)}$ are presented in Figure \ref{fig:7}. The relative differences increase toward the upper chromosphere, but remain mostly below the percent level. There is also some enhancement near shock fronts. We emphasize that these differences arise from a physical term kept in the hyperbolic treatment and are therefore primarily an error in the parabolic treatment.

Overall the hyperbolic treatment of ambipolar drift does allow for an explicit integration of the system with a time step comparable to the native MHD time step (if we include part of the corona) at minimal additional cost. For the test setups considered here with rather low field strength, the speedup compared to explicit treatment of the parabolic system is on average about $10-20$ times and scales $\sim B/\Delta x$, i.e. the benefit would be larger in active region setups or at higher resolution.

\begin{figure*}
        \centering
        \resizebox{0.95\hsize}{!}{\includegraphics{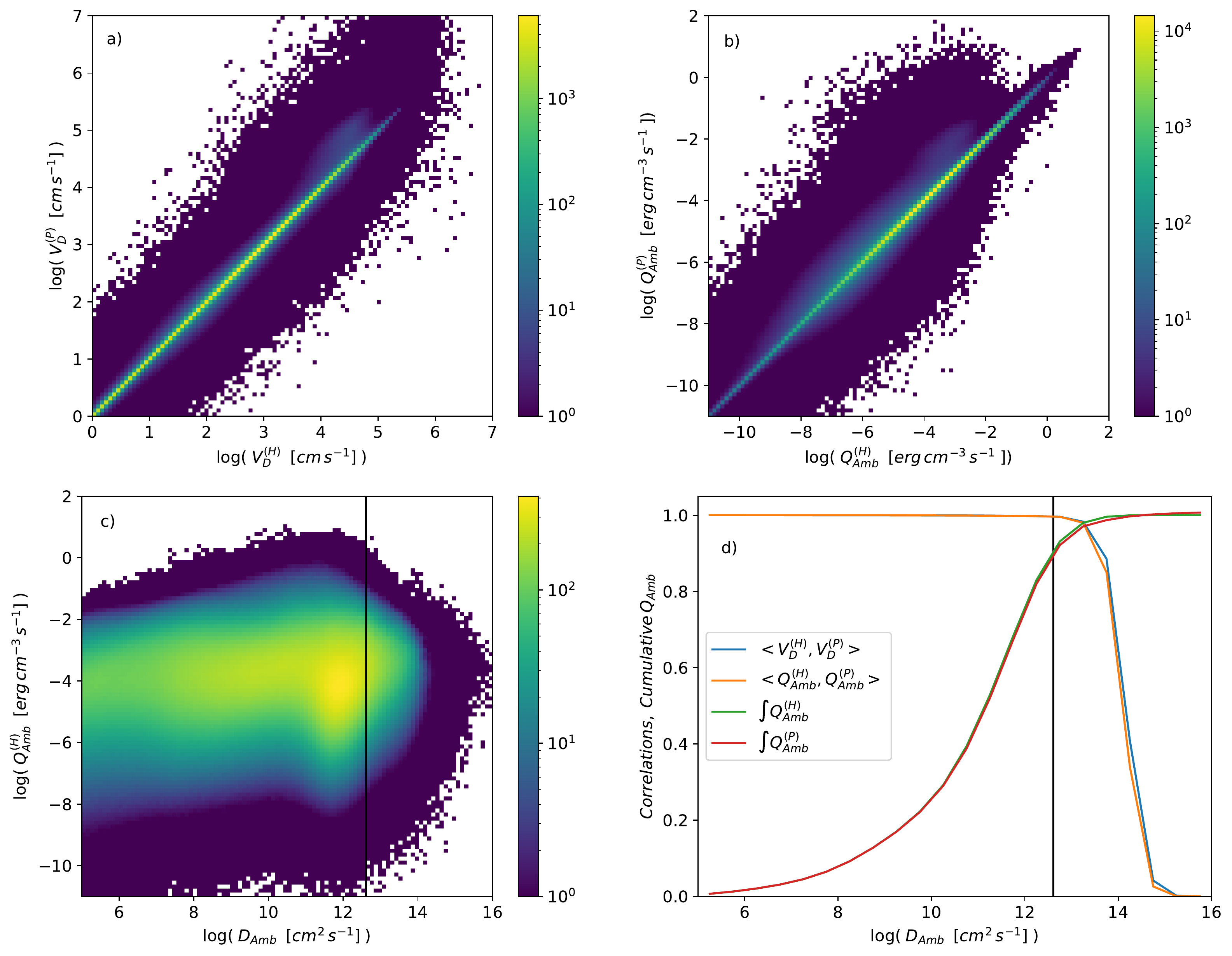}}
        \caption{Same as Figure \ref{fig:3} but for the setup using the accelerated hyperbolic treatment with 10 times larger time steps.}
        \label{fig:8}
\end{figure*}

\begin{figure*}
        \centering
        \resizebox{0.95\hsize}{!}{\includegraphics{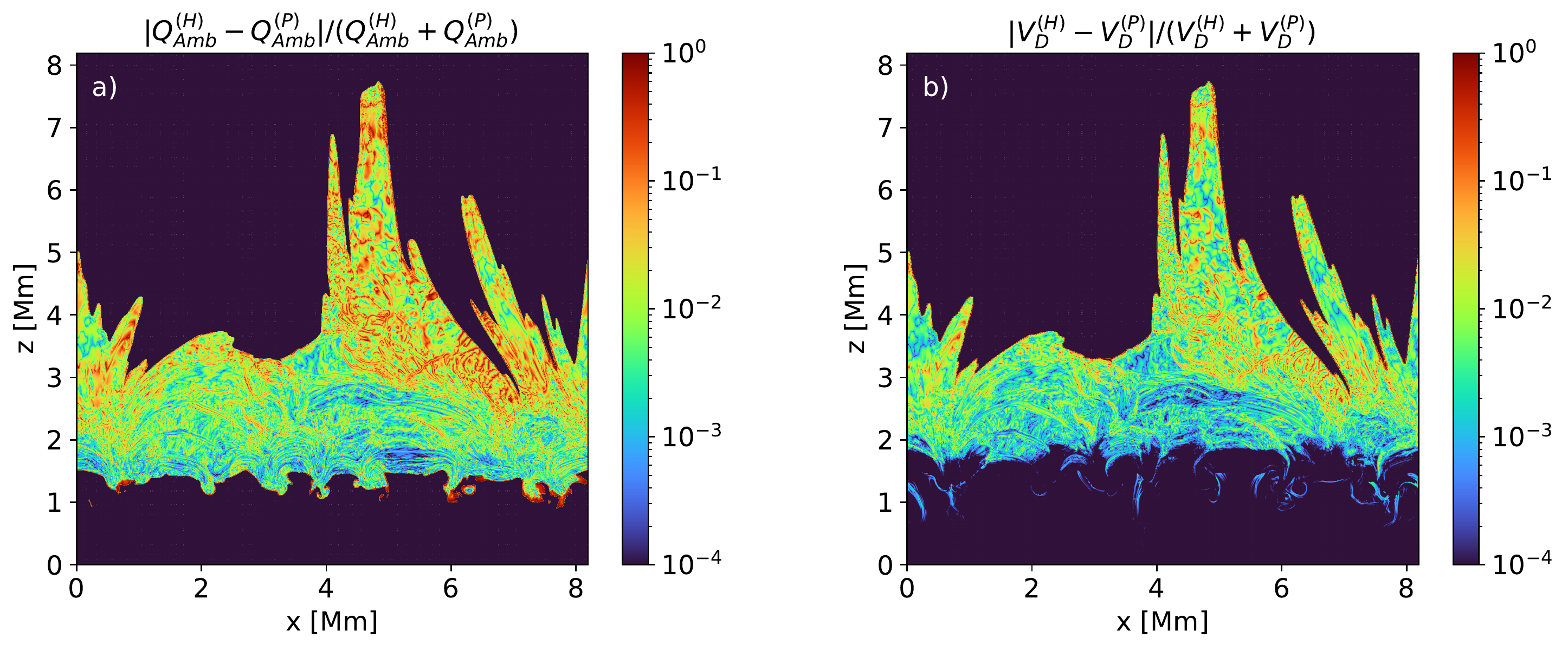}}
        \caption{Same quantities as Figure \ref{fig:7}, but for a snapshot from the simulation using the accelerated hyperbolic treatment. Relative differences reach now values that are mostly between 1\% and 10\% in the upper chromosphere. Unlike Figure \ref{fig:7} these differences are mostly errors that result from the accelerated treatment.}
        \label{fig:9}
\end{figure*} 

\subsubsection{Accelerated hyperbolic ambipolar drift}
\label{fast_hyp}
We investigate the accelerated treatment as described in Section \ref{sec:hyperbolic_acc}. To this end we use the "Boris correction" as described in \citet{Rempel:2017:corona} to artificially limit the Alfv{\'e}n velocity. The "Boris correction" is essentially semirelativistic MHD with a reduced speed of light that imposes an upper limit to the Alfv{\'e}n velocity. \citet{Rempel:2017:corona, Cheung:etal:2019:Flare} established that a choice of $c_{\rm lim}=\max(2\,C_{s\,\rm max},3\,v_{\rm max})$ (dynamically adjusted) is a good compromise between speed and the ability to capture dynamic phenomena including flares. For the rather quiescent test setup considered here, such a choice does lead to an Alfv{\'e}n cutoff velocity of about $300$ km s$^{-1}$. We use an additional safety factor of $2$, i.e. consider value of $2 c_{\rm lim}$, which allows for an integration of the system with a time step of about $0.019$ s, i.e. an additional more than tenfold speedup in integration speed compared to the hyperbolic treatment. On average this accounts for about a hundredfold increase compared to the parabolic treatment of ambipolar diffusion.

Figure \ref{fig:8} shows the same quantities as Figure \ref{fig:6} for the accelerated hyperbolic treatment. The overall level of the $V_D^{(H)}$ - $V_D^{(P)}$ and  $Q_{\rm Amb}^{(H)}$ - $Q_{\rm Amb}^{(P)}$ correlation is somewhat reduced, but remains above a $99.7\%$ level. The threshold value for $D_{\rm Amb}$ at which the hyperbolic treatment dominates is now reduced to about $4\times 10^{12} {\rm cm}^2 {\rm s}^{-1}$ from  $5\times 10^{13} {\rm cm}^2 {\rm s}^{-1}$. While the correlations $V_D^{(H)}$ - $V_D^{(P)}$ and $Q_{\rm Amb}^{(H)}$ - $Q_{\rm Amb}^{(P)}$ drop now earlier, the region with the largest contributions to the ambipolar heating remains mostly unaffected by the accelerated hyperbolic treatment.

The accelerated hyperbolic treatment of ambipolar diffusion leads to larger relative differences in  $Q_{\rm Amb}^{(H)}$ and $Q_{\rm Amb}^{(P)}$ as well as $V_D^{(H)}$ and $V_D^{(P)}$ as shown in Figure \ref{fig:9}. Unlike Figure \ref{fig:7}, where the differences were primarily an error
in the parabolic treatment, the much larger differences in Figure \ref{fig:9} are a consequence of the accelerated treatment, which is a compromise between computational speed and accuracy. Relative differences reach now values that are mostly between 1 - 10\% in the upper chromosphere.

\begin{figure*}
        \centering
        \resizebox{0.95\hsize}{!}{\includegraphics{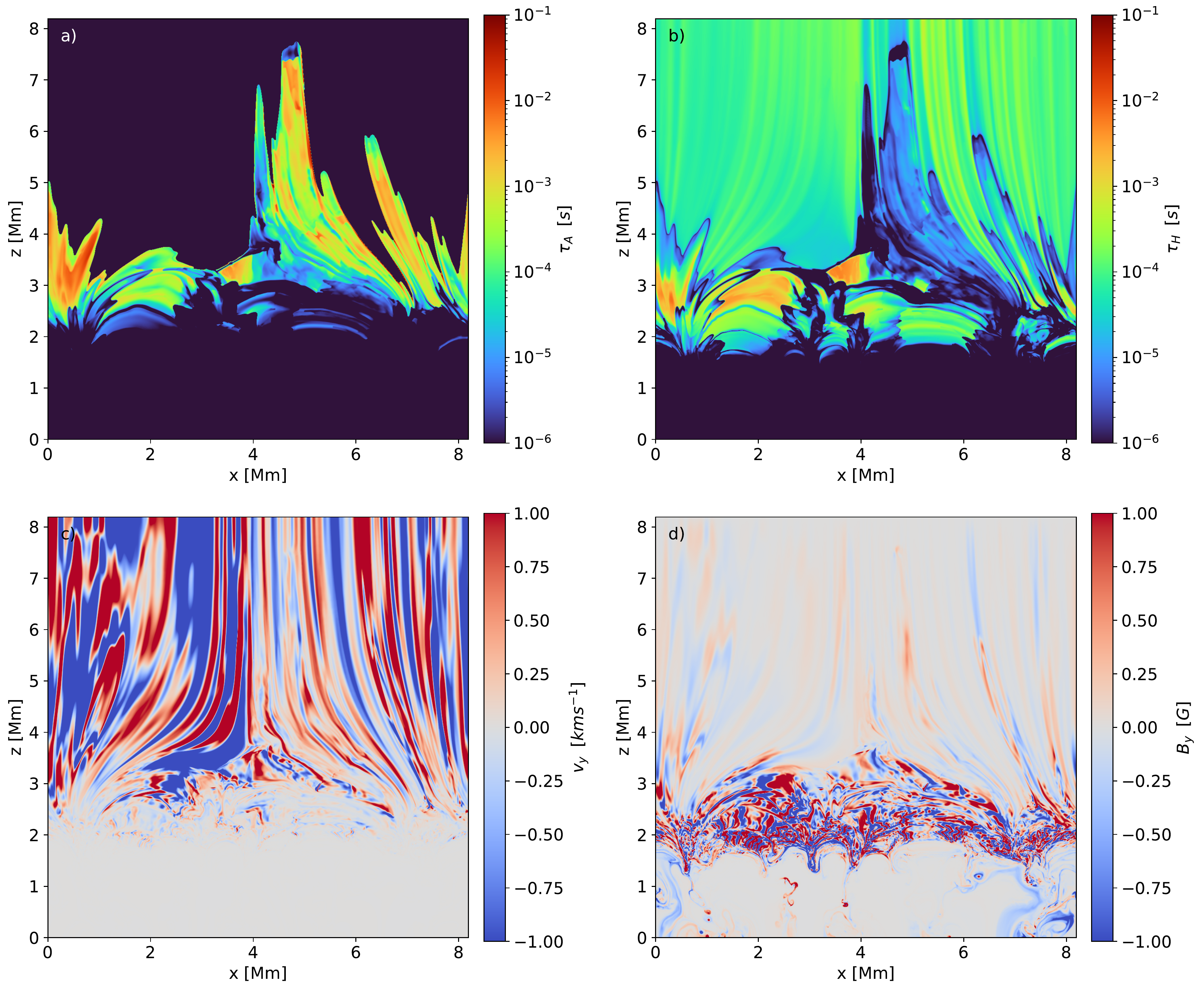}}
        \caption{Combined hyperbolic treatment of ambipolar and Hall drift. Panel (a) Relaxation time-scale $\tau_A$, (b) relaxation time-scale $\tau_H$, (c) out-of-plane ($y$) component of the flow velocity, and (d) out-of-plane ($y$) component of the magnetic 
        field.}
        \label{fig:10}
\end{figure*}

\begin{figure*}
        \centering
        \resizebox{0.95\hsize}{!}{\includegraphics{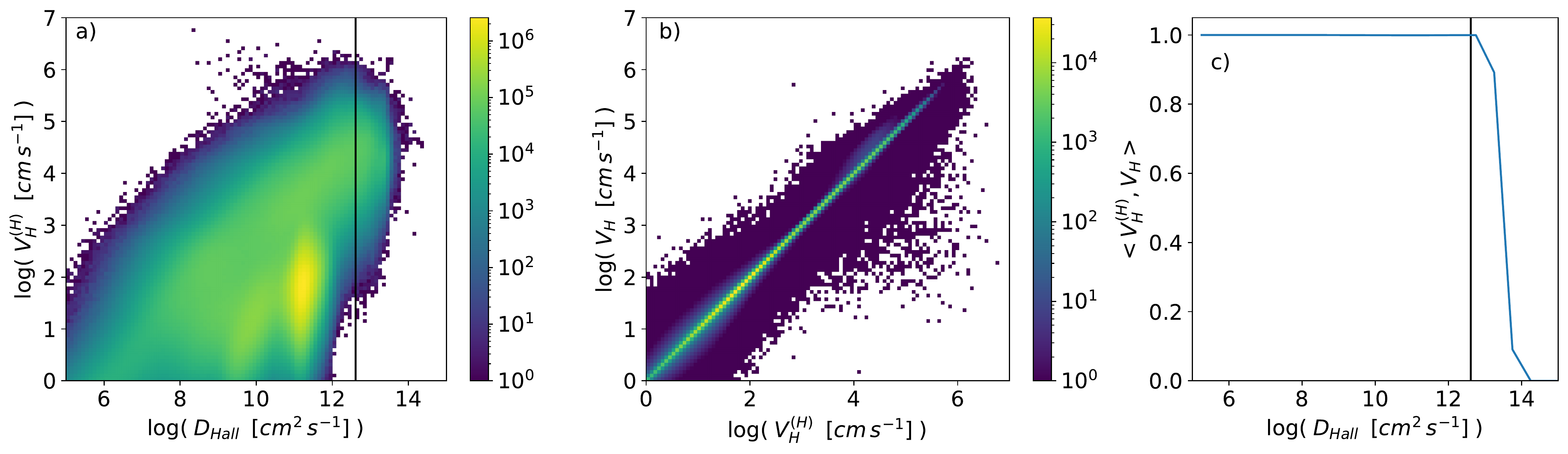}}
        \caption{Hyperbolic treatment of Hall drift. (a) 2D histogram showing the $D_{\rm Hall}$ - $V_{H}^{(H)}$ correlation,
        (b) 2D histogram showing the $V_H^{(H)}$ - $V_H$ correlation, and (c) $V_H^{(H)}$ - $V_H$ correlation as function of 
        $D_{\rm Hall}$. The vertical black line in panels (a) and (c) indicates the $D_{\rm Hall}$ value at which the hyperbolic treatment takes over.}
        \label{fig:11}
\end{figure*}

\begin{figure*}
        \centering
        \resizebox{0.475\hsize}{!}{\includegraphics{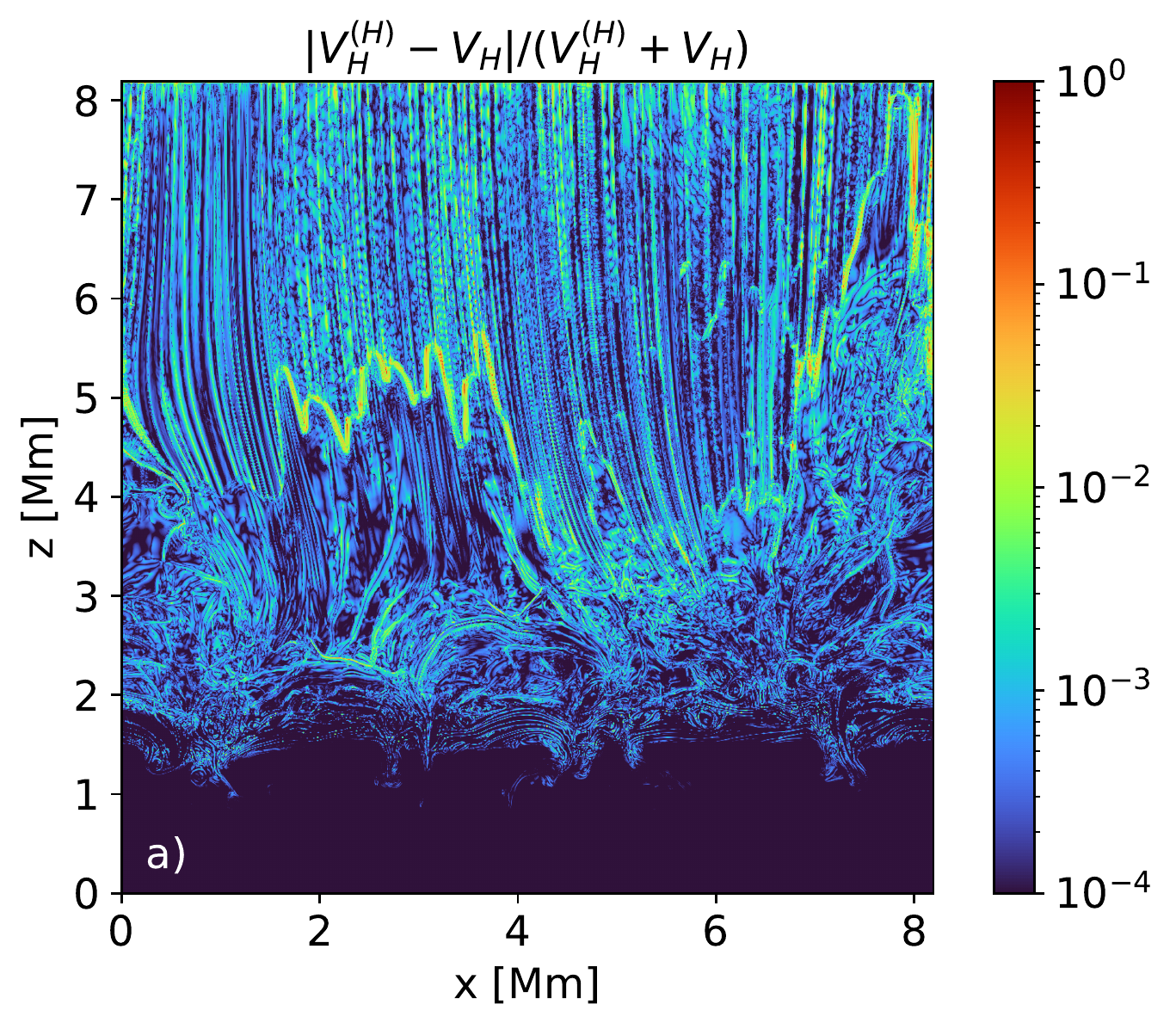}}
        \resizebox{0.475\hsize}{!}{\includegraphics{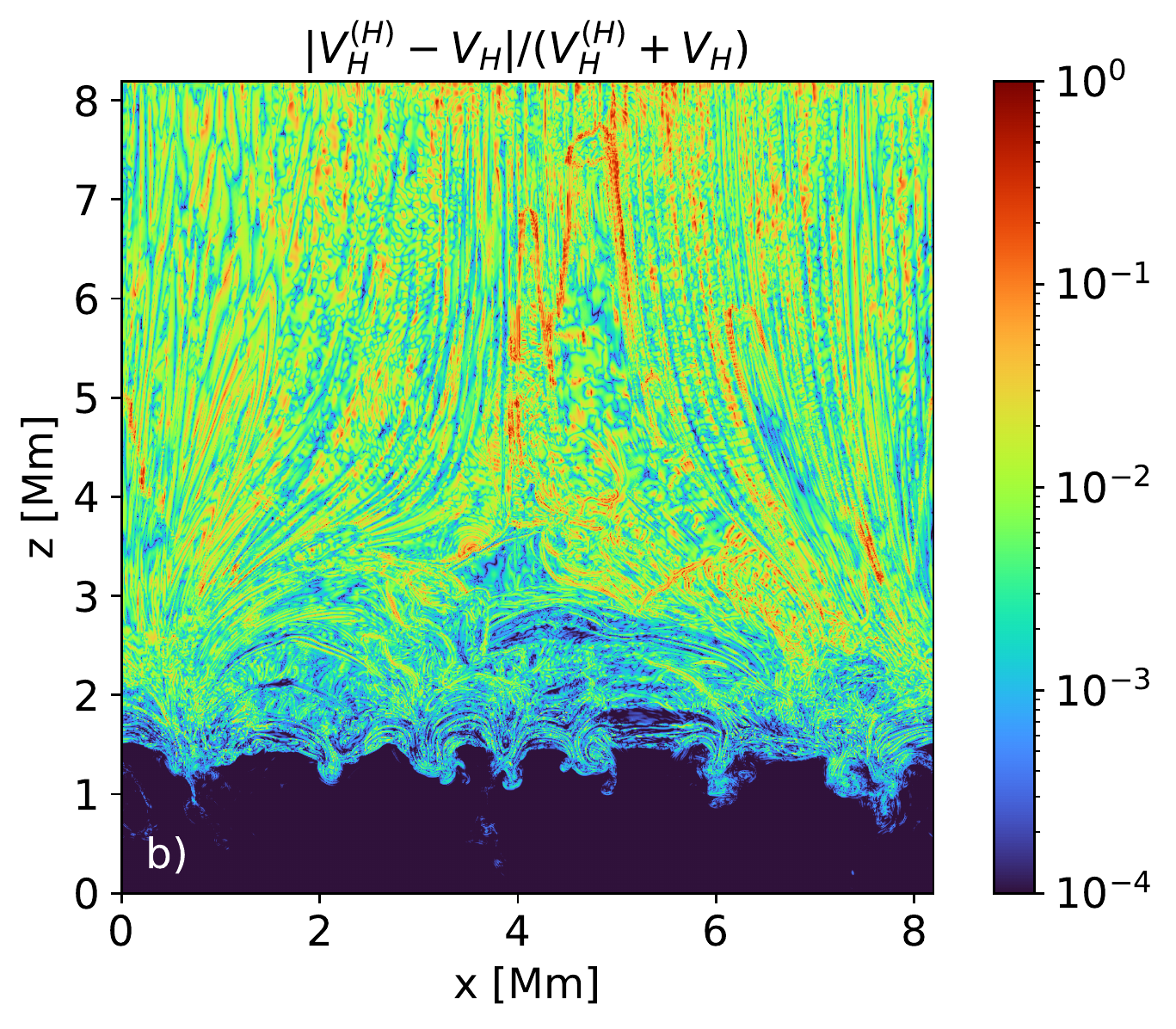}}
        \caption{Spatial distribution of relative differences between $V_H^{(H)}$ and $V_H$. Panel a) shows a snapshot from the nonaccelerated simulation presented in Figure \ref{fig:5}, panel (b) shows a snapshot from the accelerated simulation also shown in Figures \ref{fig:10} and \ref{fig:11}. The amplitude of the relative differences is comparable to those found for the ambipolar treatment presented in Figures \ref{fig:7} and \ref{fig:9}.}
        \label{fig:12}
\end{figure*}       

\subsubsection{Combined treatment of ambipolar and Hall drift}
\label{sec:test2dambhall}
Figure \ref{fig:10} shows quantities from a simulation with a combined treatment of ambipolar and Hall drift. We use here the same setup as in Section \ref{fast_hyp}. Panels (a) and (b) show the relaxation time-scales $\tau_A$ and $\tau_H$. Only physical processes that vary on timescales comparable $\tau_A$ and $\tau_H$ will be influenced by the hyperbolic treatment. Consistent with Eq. (\ref{tau_ratio}), $\tau_A$ is larger than $\tau_H$ in most locations. Panels (c) and (d) show the out-of-plane ($y$) components of velocity and magnetic field. Both are a consequence of the Hall effect and we find here amplitudes that are comparable to those reported by \citet{Cheung:Cameron:2012:muram_amb_hall} in a similar setup. While the induced Hall field is strongest in the chromosphere, the resulting Lorentz-force driven flow velocities are strongest in the coronal part of the simulation. Figure \ref{fig:11}, panel (a), presents the 2D histogram of Hall diffusivity and (hyperbolic) Hall drift velocity. The most significant contributions are found for values short of the threshold at which the hyperbolic treatment takes over (vertical black line). Consequently $V_H^{(H)}$ and $V_H$ remain strongly correlated throughout most of the domain, panels (b) and (c). Overall the accelerated hyperbolic treatment relaxed the Hall time-step constraint by about a factor of seven. In the case of the nonaccelerated hyberbolic treatment, the Hall effect is mostly unaffected except for a few rare extreme values. In this case the Hall term was not necessarily time-step limiting, but the integration as a hyperbolic system implies that the Hall term lags by about one time step behind, which leads to an error on the order of $10^{-4} - 10^{-3}$ as shown in Figure \ref{fig:12}a). In the case of the accelerated treatment we find relative errors between $V_H^{(H)}$ and $V_H$ mostly on the order of $10^{-2}$, see Figure \ref{fig:12}(b).

\begin{figure*}
        \centering
        \resizebox{0.95\hsize}{!}{\includegraphics{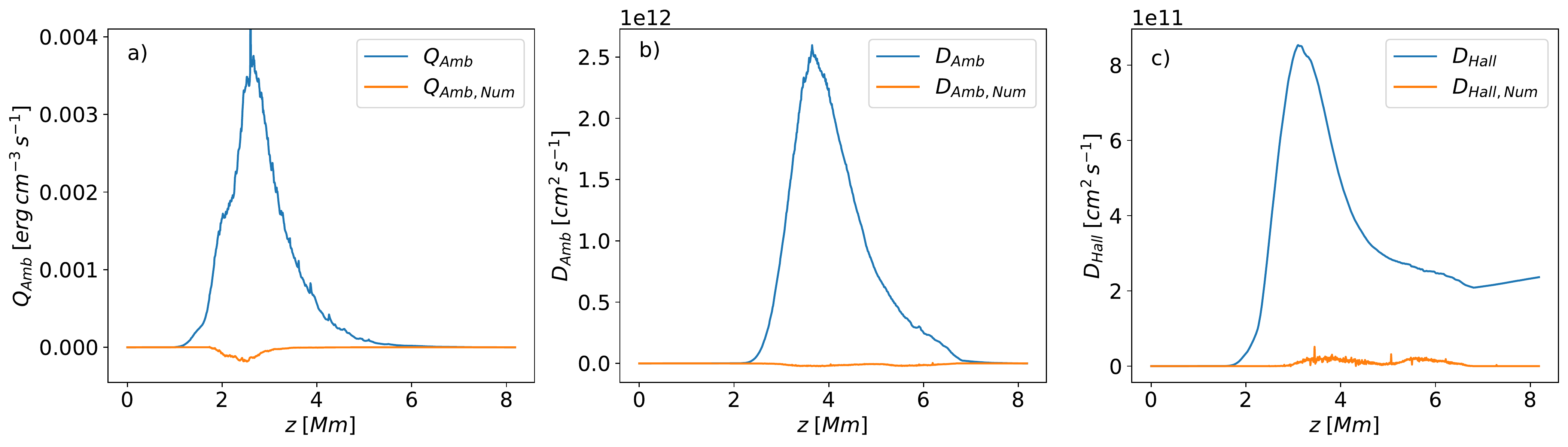}}
        \caption{Crosstalk between ambipolar and Hall diffusion resulting from hyperbolic treatment. (a) Comparison of the horizontally and temporally averaged profiles of $Q_{\rm Amb}$ (blue) and $\vec{v_H}\cdot(\vec{J}\times\vec{B})/c$ (orange). (b) Comparison of ambipolar diffusivity (blue) and the crosstalk resulting from components of $\vec{v_H}$ aligned with $\vec{J}\times\vec{B}$ (orange). (c) Comparison of Hall diffusivity (blue) and the crosstalk resulting from components of $\vec{v_D}$ aligned with $\vec{J}$ (orange).}
        \label{fig:13}
\end{figure*}

As discussed in Section \ref{sec:amb_hall}, the hyperbolic treatment does cause some crosstalk in that $\vec{v_D}$ can have $\vec{J}$ aligned components and $\vec{v_H}$ can have $\vec{J}\times\vec{B}$ aligned components. This is presented in Figure \ref{fig:13} through comparison of horizontally and temporally averaged profiles for ambipolar heating as well as ambipolar and Hall diffusivities. The analysis is performed for the simulation with accelerated treatment.
The crosstalk from $\vec{v_H}$ causes an on average negative contribution (i.e. ambipolar antidiffusivity) at a level of $-0.03$ for spurious heating and $-0.01$ for effective diffusivity. Similarly the crosstalks from $\vec{v_D}$ causes an on average positive contribution to Hall diffusivity on a $0.02$ level. In the case of the nonaccelerated treatment these values are found to be more than an order of magnitude smaller, and the contribution from $\vec{v_D}$ to Hall diffusivity is in this case physical. We do not include the contribution from  $\vec{v_H}\cdot(\vec{J}\times\vec{B})/c$ in the energy equation, since this is a purely numerical effect resulting from the hyperbolic treatment of the Hall drift. The crosstalk resulting from the accelerated treatment scales with $\tau_A$ and $\tau_H$, which scale quadratically with the achieved speedup.

\section{Conclusions}
\label{sec:concl}
We derived and implemented a set of equations that treat the effects of ambipolar drift in a hyperbolic manner. This approach differs from the classic approach of ambipolar diffusion in that the partial time derivative of the ambipolar drift velocity is retained. While this term is small and negligible in most locations, it becomes significant in those regions where the classic diffusive treatment leads to stringent time-step constraints that warrant implicit treatments or super-time-stepping approaches. We demonstrated that the hyperbolic treatment leads to a time-step constraint that is comparable to the MHD time-step constraint in a solar setup that includes the photosphere, the chromosphere, the transition region, and parts of the lower solar corona. Therefore, keeping this term is sufficient to alleviate most of the time-step constraints arising from ambipolar diffusion in a typical solar setup. More computationally expensive implicit or super-time-stepping methods are not necessary.

In addition we discussed an accelerated hyperbolic treatment that is applicable in MHD simulations that artificially limit the Alfv{\'e}n velocity through the Boris correction \citep[see, e.g.,][]{Rempel:2017:corona}. The accelerated treatment ensures that the time-step limitation from the hyperbolic treatment of ambipolar drift is not more severe than the MHD time-step constraint with a reduced Alfv{\'e}n velocity and therefore minimizes the computational expense. Since the treatment is fully explicit the direct computational overhead is small, about $15\%$ for our implementation in the MURaM code. Whereas the hyperbolic treatment with the physical collision frequency $\nu_{in}$ is physically more correct than the diffusive treatment, the accelerated hyperbolic treatment is an approximation with trade-offs between integration speed and error that have to be quantified before applying this method. Essentially the hyperbolic treatment introduces an averaging time-scale $\tau_A$ and the error introduced depends on the intrinsic time-scales of the physical problem relative to $\tau_A$.

In addition we introduced a hyperbolic treatment for the Hall drift. The hyperbolic treatment for ambipolar drift follows naturally from the multifluid equations. In the case of Hall drift, the hyperbolic nature arises from the electron momentum, which we still neglect. Instead we introduced a treatment that is formally similar to that of ambipolar drift and therefore more a numerical "trick" to avoid time-step constraints. Similar to the accelerated hyperbolic treatment this is again an approximation with trade-offs between integration speed and error, which also includes crosstalk between ambipolar and Hall diffusion. While the accelerated ambipolar treatment was found to be stable regardless of the acceleration, we did find instabilities in the case of the Hall drift, which could be controlled by enlarging the corresponding averaging time-scale in the hyperbolic treatment. In addition, larger acceleration is possible when Hall drift occurs in combination with a much larger stabilizing ambipolar drift, which is typically the case for the solar chromosphere.

The hyperbolic treatment of both ambipolar and Hall drift does cause some crosstalk between ambipolar and Hall diffusion. This was found to be on the percent level for the accelerated treatment, but more than an order of magnitude smaller for the nonaccelerated treatment.

We conducted the 2D solar tests with tabulated collision rates and electron/ion densities following from LTE. It has been found that NLTE treatment increases in general the ionization of the plasma \citep{Nobrega:2020:AmbDiffFEM} and consequently the amplitudes of ambipolar and Hall drift are reduced. It is therefore possible that the accelerated treatment suggested here will not be necessary in such simulations and the more physical hyperbolic treatment with the correct collision frequency will suffice for most chromospheric applications.

\begin{acknowledgements}
This material is based upon work supported by the National Center for Atmospheric Research, which is a major facility sponsored by the National Science Foundation under Cooperative Agreement No. 1852977. We would like to acknowledge high-performance computing support from Cheyenne (doi:10.5065/D6RX99HX) provided by NCAR's Computational and Information Systems Laboratory, sponsored by the National Science Foundation. This project has received funding from the European Research Council (ERC) under the European Union’s Horizon 2020 research and innovation program (grant agreement No. 695075). D.P. would like to thank A. Irwin (\href{https://sourceforge.net/projects/freeeos/}{Free-EoS}) for the fantastic open-source package provided and S. Danilovic, J. Martinez-Sykora, and B. De Pontieu for helpful discussion on the collision frequencies. M.R. thanks Yuhong Fan for comments on the manuscript. We thank the anonymous referee for helpful comments that improved the content and presentation of the paper.
\end{acknowledgements}

\bibliography{references}

\end{document}